\documentclass[preprint,12pt]{elsarticle}
\usepackage[T1]{fontenc}
\usepackage{amsmath,amssymb,amsthm}
\usepackage{graphicx}
\usepackage{wasysym}
\usepackage{booktabs}
\usepackage{tabularx}
\usepackage[linesnumbered,ruled,vlined]{algorithm2e}
\usepackage{tikz}
\usepackage{hyperref}
\usepackage{fancybox}
\usepackage{fancyvrb}
\usepackage[frozencache]{minted}

\usepackage{xcolor} 
\makeatletter
\let\@float@c@listing\@caption
\makeatother
\usepackage{caption}

{\bfseries}{\rmfamily}
\newtheorem{theorem}{Theorem}

\newtheorem{corollary}{Corollary}
\newtheorem{lemma}{Lemma}

\newtheorem{definition}{Definition}
\newtheorem{example}{Example}

\newcommand{\floor}[1]{\left\lfloor #1 \right\rfloor}
\newcommand{\ceil}[1]{\left\lceil #1 \right\rceil}

\usepackage{xcolor}
\definecolor{light-gray}{gray}{0.95}
\newcommand{\code}[1]{\colorbox{light-gray}{\texttt{#1}}}
\usepackage{array}
\newcolumntype{L}{>{$}l<{$}}
\newcommand{\sepa}{\kern0.25em\textbullet\kern0.25em}

\newcommand{\cert}{\mathtt{cert}}
\newcommand{\asm}{\mathtt{asm}}
\newcommand{\rnd}{\mathtt{rnd}}
\newcommand{\lin}{\mathtt{lin}}
\newcommand{\uns}{\mathtt{uns}}
\newcommand{\sol}{\mathtt{sol}}
\newcommand{\reason}{\mathtt{reason}}
\newcommand{\data}{\mathtt{data}}
\newcommand{\infeasible}{\mathtt{infeasible}}
\newcommand{\R}{\mathbb{R}}
\newcommand{\Z}{\mathbb{Z}}
\newcommand{\A}{\mathtt{A}}
\newcommand{\formula}{\phi}
\newcommand{\OR}{\vee}
\newcommand{\AND}{\wedge}
\newcommand{\bigOr}{\bigvee}
\newcommand{\bigAnd}{\bigwedge}
\newcommand{\true}{\mathtt{true}}
\newcommand{\false}{\mathtt{false}}
\newcommand{\ttif}{\mathtt{\; if \;}}
\newcommand{\ttelse}{\mathtt{\; else \;}}
\newcommand{\ttthen}{\mathtt{\; then \;}}

\newcommand{\nz}{\mathtt{nz}}
\newcommand{\sense}{\mathtt{sense}}
\newcommand{\NULL}{\mathtt{null}}

\usepackage{subfiles}

\title{Satisfiability Modulo Theories for Verifying MILP Certificates}
\author[1]{Kenan Wood\corref{cor1}}
\author[2]{Runtian Zhou}
\author[3]{Haoze Wu}
\author[4]{Hammurabi Mendes}
\author[5]{Jonad Pulaj}
\address[1]{Davidson College, 225 Baker Dr, Box 5000, Davidson, NC 28035; kewood@davidson.edu}
\address[2]{Duke University, Durham, NC 27708; daniel.zhou@duke.edu}
\address[3]{Amherst College, Amherst, MA 01002; hwu@amherst.edu}
\address[4]{Davidson College, Mathematics and Computer Science, 209 Ridge Rd, Box 5000, Davidson, NC 28035; hamendes@davidson.edu}
\address[5]{Davidson College, Mathematics and Computer Science, 209 Ridge Rd, Box 5000, Davidson, NC 28035; jopulaj@davidson.edu}
\cortext[cor1]{Corresponding author. Email: kewood@davidson.edu}

\begin{document}

\begin{abstract}
Correctness of results from mixed-integer linear programming (MILP) solvers is critical, particularly in the context of applications such as hardware verification, compiler optimization, or machine-assisted theorem proving. To this end, VIPR 1.0 is the first recently proposed general certificate format for answers produced by MILP solvers.
We design a schema to encode VIPR's inference rules as a ground formula that completely characterizes the validity of the algorithmic check, removing any ambiguities and imprecisions present in the specification. 
We formally verify the correctness of our schema at the logical level using \texttt{Why3}'s automated deductive logic framework.
Furthermore, we implement a checker for VIPR certificates by expressing our formally verified ground formula with the Satisfiability Modulo Theory Library (SMT-LIB) and check its validity. Our approach is solver-agnostic, and we test its viability using benchmark instances found in the literature.

\end{abstract}


\maketitle

\section{Introduction}
\label{sec:intro}

Computational mixed-integer linear programming (MILP) is widely considered a successful blend of algorithmic improvement with advancements in hardware and compilers~\cite{bixby2012brief}, where current state-of-the-art MILP solvers can solve problems with millions of variables and constraints~\cite{koch2022progress}. 

Demand for MILP solvers is largely driven by applications in industry, hence the underlying numerical computations typically rely on floating-point arithmetic for efficiency. For most such applications, in order to achieve a high degree of numerical stability, floating point computations are combined with numerical error tolerances. 
However, MILP solvers are also used in experimental mathematics to find counterexamples~\cite{graphtheory,pulajfrankl1} or provide numerical evidence~\cite{pebbling,pulaj2023local,ManickamMiklosSinghi}. In such theorem proving applications, correctness is tantamount and additional safeguards are needed to counter the use of inexact floating-point arithmetic and/or programming or algorithmic errors. The best known example in this case is the verification in \texttt{Isabelle/HOL}~\cite{10.5555/1791547} of the infeasibility of thousands of linear programs involved in the proof of Kepler's conjecture \cite{kepler}.  

Another set of tools that have been successfully deployed in automated theorem proving tasks are  Boolean Satisfiability (SAT) solvers and Satisfiability Modulo Theories (SMT) solvers~\cite{smt}. For example, SAT solvers have had success in settling a number of long-standing mathematical problems including the Erd\"os discrepancy conjecture~\cite{konev2015computer}, the Boolean Pythagorean triples conjecture~\cite{drat}, and the determination of the fifth Schur number~\cite{heule2018schur}. On the other hand, SMT solvers are increasingly used as subroutines for proof assistants~\cite{bohme2010sledgehammer,ekici2017smtcoq} to automatically resolve proof goals. In those applications, SAT/SMT solvers are often required to produce proofs, which can be independently checked by proof checkers to guarantee the correctness of the solver outputs. One theory particularly relevant to MILP is the theory of Linear Integer Real Arithmetic, for which proof production is supported in state-of-the-art SMT solvers such as \texttt{Z3}~\cite{z3}, \texttt{cvc5}~\cite{barbosa2022cvc5}, and \texttt{MathSAT5}~\cite{10.1007/978-3-642-36742-7_7}.

Although verification of answers returned by MILP solvers is not as well-established as in analogous solvers in the SAT/SMT community\footnote{For analogous efforts in the Constraint Programming community we refer the reader to the following works~\cite{gocht2022auditable, demirovic2024pseudo}.}, significant steps forward have been made. The first efforts in this direction were the certification of the optimality of a large TSP problem~\cite{APPLEGATE200911}, followed by the use of subadditive generative functions, which involve \emph{finding} solutions of ``dual'' MILPs, for the certification of optimality~\cite{subadditivefunct} (but not infeasibility). An important milestone is the proposal of \texttt{VIPR 1.0} (Verifying Integer Programming Results), the first branch-and-cut \emph{general} certificate format for MILP instances which handles optimality, infeasibility, and relaxed optimality bounds (see Theorem~\ref{thm-vipr-validity} for a precise statement). \texttt{VIPR 1.0} is designed with simplicity in mind, and is composed of a list of statements that can be sequentially verified using a limited number of inference rules \cite{vipr1,vipr2}. This certificate format, with additional safeguards, was used in the proof of special cases of Chv{\'a}tal's conjecture \cite{vipr3}, and in the proof of the 3-sets conjecture~\cite{pulajfrankl2}. The \texttt{VIPR 1.0} format\footnote{\texttt{VIPR 1.1}~\cite{vipr2} extends inference rules derived with floating point arithmetic. These are then transformed into \texttt{VIPR 1.0} format and verified with \texttt{viprchk}.} is currently implemented in the latest versions of \texttt{exact SCIP}\footnote{Currently state-of-the-art hybrid-precision implementation for solving MILP instances exactly over rational numbers.}~\cite{cook2011exact}, together with a checker (\texttt{viprchk} implemented in C++) for the corresponding certificates. For brevity, in the rest of the paper we refer to \texttt{VIPR 1.0} as \texttt{VIPR}; mathematical formalisms of the \texttt{VIPR} format in this paper drop the teletype font (e.g. we formally discuss VIPR certificates, but our implementation uses the \texttt{VIPR} file format).

While \texttt{viprchk} remains useful as a reference and is geared toward performance, it notably lacks formal reasoning associated with its algorithmic correctness. To further complicate matters, some \texttt{VIPR} specifications are informally worded, or not explicitly stated as we point out in Subsection~\ref{subsec-differences}. Additionally, the implementation of \texttt{viprchk} is \emph{not} entirely MILP solver agnostic: depending on the certificate it is examining, \texttt{viprchk} may rely on \emph{expected} MILP solver behavior for its algorithmic check instead of strictly relying on specifications. To illustrate this, in Subsection~\ref{subsec-examples} and \ref{App-ForgedManipuated} we exhibit simple VIPR certificates for which \texttt{viprchk} returns incorrect or ambiguous answers. 
 Although it is unlikely an established MILP solver produces such certificates or similar ones, it is nevertheless possible given the specifications. Finally in Subsection~\ref{subsec-differences} we point out how \texttt{viprchk}'s non-skeptical use of the \texttt{VIPR} certificate's \texttt{index} attribute, which aims at memory management and fast performance, fundamentally requires trusting the associated MILP solver at the potential expense of verification correctness. 

Our motivation for this work is addressing the issues outlined above in order to enable viable alternatives to \texttt{viprchk}. Our main contributions are as follows. First, since some \texttt{VIPR} specifications are described in informal language, we remove possible ambiguity by introducing more formal definitions which lead to a \emph{precise} notion of validity (Section~\ref{sec:vipr}), which in particular does not make use of the \texttt{index} attribute. Second, we formalize the equivalence of a VIPR certificate validity with a variable-free quantifier-free predicate (ground formula) in the theory
of Linear Integer Real Arithmetic (Theorem~\ref{thm-equivalence}).
Third, we formally verify the correctness of our predicate construction with the deductive logic framework of \texttt{Why3} \cite{Why3}, showing that a VIPR certificate is valid if and only if the corresponding predicate evaluates to true.
Finally, we 
implement a checker framework by expressing the equivalence predicate using the Satisfiability Modulo Theory Library (SMT-LIB) and check its validity with \texttt{cvc5}. In Section~\ref{sec:expts} we evaluate our checker using relevant benchmark instances from the literature.

There are advantages to our approach beyond verification, as our formalism helps shed light on the desired structure of the specifications themselves. For example, we see in Subsection~\ref{sec:permissive} that some \texttt{VIPR} specifications are \emph{permissive} in the sense that they allow vacuously valid derivations that a MILP solver would not consider in the first place. While this permissive notion of validity can be desirable in some contexts, it can impact verification time with redundant or useless information. Our formalism can be adapted to both stronger and weaker notions of validity, depending on application goals, with minimal changes to the specifications and to our equivalency predicate. Finally, additional inference rules can easily be integrated into our existing framework given its flexibility.

Our contributions towards verification are two-fold. Firstly, we formally verify the correctness of our transformation \textit{at the logical level} using \texttt{Why3}'s automated deductive logic framework \cite{Why3}. The main advantage of this is that our transformation is \textit{implementation-independent}. Depending on the desired level of trust of runtime computations, one may easily evaluate our predicate using any tool of choice, since the predicate is written completely unambiguously and is formally verified.

Secondly, we implement our construction using \texttt{cvc5}, a state-of-the-art SMT solver, ensuring robust verification at runtime in addition to at the logical level.
However, we stress that our implemented checker is SMT solver \emph{agnostic}, as \emph{any} SMT-LIB compliant solver can be used to evaluate the equivalence predicate with small changes to our implementation. Furthermore, since we are evaluating a ground formula, our use of the SMT solver \texttt{cvc5} in Section~\ref{sec:expts} can instead be replaced with model validation tools like Dolmen~\cite{bury2023verifying}. Importantly, the Boolean expressions of our conjunctive formula easily allow for parallelism in their checking, a feature that we explore in our analytical framework. Finally with ongoing projects like \texttt{SMT Lean}~\cite{githubGitHubUfmgsmiteleansmt}, which support the theories of Linear Integer Real Arithmetic, further integration with an interactive theorem prover is within reach with additional work. We refer the reader to Section~\ref{sec:further_integ} for further discussions on moving towards a safer checker for VIPR certificates and potential integrations or further extensions. All associated code with this project is freely available on GitHub.\footnote{\href{https://github.com/hammurabi-mendes/vipr_checker}{https://github.com/hammurabi-mendes/vipr\_checker}, \href{https://github.com/JonadPulaj/vipr-gator/tree/main}{Why3 formalization}}
The rest of this paper is organized as follows. Section \ref{sec:vipr} provides an overview of MILP and VIPR certificates, which we write abstractly for simplicity, but still without loss of generality; Section \ref{sec:vipr} also characterizes the validity of a given VIPR certificate. Section \ref{sec:smt} carefully defines a ground formula from a VIPR certificate, and further proves the equivalence of the formulation. Section \ref{sec:expts} presents an evaluation of our design tested on known benchmarks. Finally, our concluding remarks are in Section \ref{sec:concl}.

\section{MILP Branch-and-Bound Background}\label{Sec:MILP-branch-and-bound}
Before formally describing VIPR certificates in a rigorous mathematical way, we first give some necessary background on MILP and branch-and-bound techniques for MILP solving. 
First, we describe an informal overview of MILP and branch-and-bound techniques used in \texttt{exact SCIP}.
In order to safe-guard against computational instability and potential bugs, \texttt{exact SCIP} uses a small set of procedures that perform well in practice and ensure robust verifiability. This is in contrast to other floating-point commercial solvers \cite{gurobi,cplex2009v12} that use more complex procedures to improve performance.

A MILP contains of a set of \emph{variables}, some taking on real values (called \emph{real variables}, and some restricted to integer values (called \emph{integer variables}). There is also a linear \emph{objective function} to optimize, along with an indication of \emph{maximization} or \emph{minimization}, describing the direction the objective function is being optimized. Finally, there is also a set of linear \emph{constraints} (equalities or inequalities), which constrain the possible \emph{feasible} values of the variables to satisfy every constraint. Then, the problem of the MILP is to optimize the objective function over all feasible solutions (those that satisfy the constraints and integrality requirements of the integer variables).

There are a variety of operations and insights that are fundamental to branch-and-bound methods for MILP solving. First, if there is a constraint that is implied by the starting constraints (described by the following operations, formally described in Definition \ref{def-derived-constraint}) and is trivially always false (Definition \ref{def-absurdity}), such as $0 \le -1$, then the MILP is clearly infeasible, meaning that no feasible solutions exist. This is the notion of an \emph{absurdity} in Definition \ref{def-absurdity}.
If we have established a set of constraints that are implied by the original ones, then we can obtain a new constraint by taking an appropriate linear combination (one that is compatible with the directions of the inequalities). This is the notion of a \emph{suitable linear combination} in Definition \ref{def-linearcomb}. Similarly, if the linear combination of variables in an inequality constraint (its left hand side) is always an integer, as determined by the integer variables, then the constant part of the inequality (the right hand side) can safely be \emph{rounded} to an integer without changing the set of feasible solutions; see Definition \ref{def-rounding}. Additionally, if we obtain a feasible solution, then this naturally gives a bound that all optimal solutions must satisfy, if an optimal solution exists; see the \texttt{sol} reasoning in Definitions \ref{def-derived-constraint} and \ref{def-valid-der-constraint}.

The most fundamental idea for MILP solving is that of the \emph{split disjunction}, and is the ``branching'' part of branch-and-bound. At any stage of solving, it is often helpful to artificially ``add in'' certain constraints, which can be used with suitable linear combinations later into solving. In certain cases, we can split the feasible space of solutions into two, each lying in one-half of the space. For example, if $x_1$ and $x_2$ are integral and $x_3$ is real, then every feasible solution must satisfy either $x_1 + 2x_2 \ge 3$ or $x_1 + 2x_2 \le 2$, since the left-hand-side of both of these inequalities must be an integer. These two inequalities form a \emph{split disjunction}, as in Definition \ref{def-disjunction}. Thus, we can add in these inequalities by instantiating two distinct ``copies'' of the MILP, one with the extra constraint $x_1 + 2x_2 \ge 3$, and one with $x_1 + 2x_2 \le 2$. This creates the initial two branches of the tree, which can be branched further at any time when useful. Once both branches have been solved (using a recursive-style branching algorithm), the results can be appropriately aggregated together. This yields either infeasibility if both branches are infeasible, or that the optimal solution is the best of the two branches. Individual constraints can also be propagated up the tree in this way.

Using this framework, VIPR certificates essentially ``replay'' the generation of each node (a constraint) of the branch-and-bound tree, indicating every suitable linear combination, rounding, and feasible solution constraints. Note that the certificate gives the necessary data to generate each constraint (found by the solver), so the checker can independently replay the solving process, in the sense that replay is checking the correctness of a candidate witness. Furthermore, the branch-and-bound tree is in a sense ``flattened'' through the notion of \emph{assumptions}, which is the set of constraints obtained through split disjunctions (formalized by \texttt{asm} reasoning for VIPR derived constraints). For every constraint derived, we indicate the branch of the tree that it corresponds to by providing the set of assumption constraints required to generate it. If a derived constraint holds in two different branches, then we can \emph{unsplit} the branches, upward propagating the constraint; this reasoning is formalized with the \texttt{uns} keyword in VIPR.

\subsection{MILP Formalism}\label{Subsec:MILP-formalized}
Now let us present some formal notions and definitions for MILP.
We note that all real values (coefficients, solutions, linear combination multipliers, etc.) discussed below are assumed to be appropriately rational or integral in a VIPR certificate for arithmetic exactness. In particular, our checker assumes that rational inputs are expressed as $a/b$ for appropriate integers $a$ and $b$, and not expressed in decimal notation.

Let $x_1, \dots, x_n$ be a list of real and integer variables. Define $x \triangleq (x_1, \dots, x_n)$ as a vector in $\R^n$. Given vectors $a,b \in \R^n$, let $a \cdot b$ denote the \emph{dot product} (standard inner product) of $a$ and $b$. A \emph{(linear) constraint} on $x_1, \dots, x_n$ is an expression of the form $a\cdot x = b$, $a\cdot x \ge b$, or $a \cdot x \le b$, where $a \in \R^n$ and $b \in \R$; the \emph{left hand side} of all these constraints is the linear functional $a\cdot x$, and the \emph{right hand side} is the real number $b$. The left hand side of a constraint $C$ is denoted $l(C)$, and the right hand side of $C$ is denoted $r(C)$.
Let $C_1, \dots, C_m$ be constraints on $x_1, \dots, x_n$; let $c \in \R^n$; let $I$ be the set of $i \in [n]$ such that $x_i$ is an integer variable. Then the problem $\max\{c \cdot x: x \text{ satisfies } C_1, \dots, C_m \text{ and } \forall i\in I, x_i \in \Z\}$ is called a \emph{maximization mixed integer linear program} (MILP), and $\max\{c \cdot x: x \text{ satisfies } C_1, \dots, C_m \text{ and } \forall i\in I, x_i \in \Z\}$ is called a \emph{minimization MILP}. The \emph{constraints} of both of these optimization problems are the list $C_1, \dots, C_m$. The \emph{objective function} of these MILPs is the linear functional $c \cdot x$. Let $IP$ be a maximization or minimization MILP as described above, and define
{\footnotesize
\[
\sense(IP) \triangleq \begin{cases}
    \mathtt{max}, & \text{if $IP$ is a maximization MILP}\\
    \mathtt{min}, & \text{if $IP$ is a minimization MILP}.
\end{cases}
\]
}

A point $s \in \R^n$ is a \emph{feasible solution} of $IP$ if $s$ satisfies all the constraints of $IP$ and $s_j \in \Z$ for all $j \in I$. A feasible solution $s$ is called \emph{optimal} if $s$ minimizes (resp. maximizes) the objective value over all feasible solutions, when $\sense(IP) = \mathtt{min}$ (resp. $\sense(IP) = \mathtt{max}$). We say $IP$ is \emph{feasible} if $IP$ has a feasible solution, and \emph{infeasible} otherwise. Similarly, $IP$ is \emph{optimal} if $IP$ has an optimal solution. For any integer $n \ge 1$, define $[n] \triangleq \{1, \dots, n\}$.

\begin{example}\label{example-running-IP}
    We make use of the following running example of a MILP throughout this paper, denoted $IP_0$. This is an example of an infeasible MILP, so we omit the objective function.
    \begin{align*}
        C_1: & 2x_1 + 3x_2 \ge 1\\
        C_2: & 3x_1 - 4x_2 \le 2\\
        C_3: & -x_1 +6x_2 \le 3\\
        & x_1, x_2 \in \Z.
    \end{align*}
    In Example \ref{example-cert}, we will describe a formal certificate that $IP_0$ is infeasible; furthermore, we illustrate in Example \ref{example-cert-valid} that the certificate is valid with respect to Definition \ref{def-vipr-valid}. Both this example MILP and its certificate of infeasibility are taken from \cite{vipr1}. \hfill $\square$
\end{example}

Definitions \ref{def-sign} through \ref{def-linearcomb}, with minor differences in notation, can be found in the technical specifications of the original \texttt{VIPR} certificate \cite{GitHub-Ambros,vipr1}; the definitions in Section \ref{sec:vipr} are novel.
For brevity we simply refer to constraints without specifying its variables when convenient; furthermore, the underlying list of variables $x_1, \dots, x_n$ as well as the context of a given $IP$ is implied when not explicitly stated.
\begin{definition}[Sign]\label{def-sign}
    Let $C$ be a constraint. Define the \emph{sign} of $C$ by
    {\footnotesize
    \[
    s(C) \triangleq \begin{cases}
        1 & \text{if $C$ is a $\ge$ constraint}\\
        0 & \text{if $C$ is a $=$ constraint}\\
        -1 & \text{if $C$ is a $\le$ constraint}.
    \end{cases}%
    \]}
\end{definition}

In the following, an \emph{absurdity} is a linear constraint that is trivially false.
\begin{definition}[Absurdity]\label{def-absurdity}
   Let $\beta \in \R$ such that $\beta \neq 0$. A constraint of the form $0 \ge \beta$ for $\beta > 0$ or $0 \le \beta$ for $\beta < 0$ is called an \emph{absurdity}.
\end{definition}

In the following definition, we may replace the word ``dominate'' with ``imply'' for intuition.
\begin{definition}[Domination]\label{def-domination}
    Suppose $a \in \R^n$ and $b \ge b'$. Then the constraints $a \cdot x \ge b$ and $a \cdot x = b$ both \emph{dominate} $a \cdot x \ge b'$, the constraints $a \cdot x \le b'$ and $a \cdot x = b'$ both \emph{dominate} $a \cdot x \le b$, and finally $a \cdot x = b$ \emph{dominates} $a \cdot x = b$. An absurdity \emph{dominates} every constraint.
\end{definition}

In the following definition, two constraints are said to form a \emph{split disjunction} if every point in $\R^n$ satisfying the integrality requirements on the integer variables must satisfy exactly one of the two given constraints.
\begin{definition}[Split Disjunction]\label{def-disjunction}
    An unordered pair of two constraints $a \cdot x \le \delta$ and $a \cdot x \ge \delta + 1$ forms a \emph{split disjunction} if $\delta \in \Z$, $a \in \Z^n$ and $a_i = 0$ for all $i\in [n] \setminus I$.
\end{definition}

For certain constraints, we can guarantee that all feasible points result in an integral left hand side, which allows us to appropriately round the right hand side to an integer as well without shrinking the feasible space.
\begin{definition}[Rounding]\label{def-rounding}
    Let $C$ be a constraint with $l(C) = \sum_{i=1}^n a_ix_i$ such that $a_j \in \Z$ for all $j \in I$ and $a_j = 0$ for all $j \notin I$, $r(C) = b$, and nonzero sign. Then $C$ is called \emph{roundable}. If $C$ is a $\le$ (resp. $\ge$) constraint, then $\sum_{i=1}^n a_ix_i \le \floor{b}$ (resp. $\sum_{i=1}^n a_ix_i \ge \ceil{b}$) is called the \emph{rounding of $C$}, denoted by $\rnd(C)$.
\end{definition}

When the signs of a list of constraints ``match up'' with a list of corresponding weights, we may take linear combinations.
\begin{definition}[Suitable Linear Combination]\label{def-linearcomb}
    Consider a list of constraints $C_1, \dots, C_k$. Let $\lambda \in \R^k$. If either $\lambda_i s(C_i) \ge 0$ for all $i \in [k]$ or $\lambda_i s(C_i) \le 0$ for all $i \in [k]$, the point $\lambda$ is said to \emph{generate a suitable linear combination} of $C_1, \dots, C_k$. The \emph{resulting constraint} $C$, often denoted $\sum_{i=1}^k \lambda_i C_i$, is such that $l(C) = \sum_{i=1}^k \lambda_i l(C_i)$, $r(C) = \sum_{i=1}^k \lambda_i r(C_i)$, and sign
    {\footnotesize
    \[
    s(C) = \begin{cases}
        1 & \text{if } (\forall i \in [k], \lambda_i s(C_i) \ge 0) \AND (\exists i \in [k], \lambda_i s(C_i) > 0)\\
        0 & \text{if } \forall i \in [k], \lambda_i s(C_i) = 0\\
        -1 & \text{if } (\forall i \in [k], \lambda_i s(C_i) \le 0) \AND (\exists i \in [k], \lambda_i s(C_i) < 0).
    \end{cases}
    \]
    }%
\end{definition}

\section{Validity of VIPR Certificates}
\label{sec:vipr}

In this section we define a VIPR certificate and characterize its validity. We abstract away some technical specifications to focus on desired mathematical properties without loss of generality, and discuss such differences in Subsection~\ref{subsec-differences}. Note that all of the definitions in this section are novel.

We will find the following notation---of the \emph{non-zero index set} of a list of real numbers---helpful throughout this and the following section. This definition is only used for semantic convenience, and not necessary for the formalization.


\begin{definition}[Non-zero index set]\label{def-indexed-point}
    For a list of real numbers $p \in \R^d$, define $\nz(p)\triangleq \{i \in [d]: p_i \ne 0\}$.
\end{definition}

Now we introduce the derived constraint format used in VIPR certificates. Note that the definition is merely a constraint with extra attributes that will only make sense in the context of a VIPR certificate described below. A derived constraint is a constraint $C$ in a VIPR certificate that also specifies \textit{the method} in which it is generated ($\reason(C)$), the data inputs to this algorithm ($\data(C)$), and the assumption constraints its validity depends on ($\A(C)$). We discuss the meaning of each of these reasoning and data pairs right after Definition \ref{def-vipr}.

\begin{definition}[Derived Constraint]\label{def-derived-constraint} 
    A \emph{derived constraint} is a constraint $C$ with additional attributes $\reason(C) \in \{\asm,$ $\lin, \rnd, \uns, \sol\}$, $\data(C)$ (type depending on $\reason$), and $\A(C) \subseteq \Z_{>0}$.
    The keyword $\reason(C)$ is called the \emph{reasoning} of $C$; $\data(C)$ is called the \emph{data} of $C$; and elements of $\A(C)$ are called \emph{assumptions} of $C$.
\end{definition}

Informally, a VIPR certificate for a MILP $IP$ consists of the following: a \textit{relation to prove} (RTP), which is either infeasibility or bounds on optimality; a \textit{solution set} (SOL), which consists of a set of points to check for feasibility and satisfaction of an optimality goal (if RTP is not infeasible); and finally a list of \textit{derived constraints} (DER) that are each implied by previous constraints and by assumptions in the assumption set. For readability, note that the derived constraint reasoning strings $(\asm, \lin, \rnd, \uns, \sol)$ represent (assumption, linear combination, rounding, unsplit, solution).

We also abuse notation slightly by writing $[-\infty, \infty] \triangleq \R$ and $[-\infty, a] \triangleq (-\infty, a]$ and $[a, \infty] \triangleq [a, \infty)$ for any $a \in \R$.

\begin{definition}[VIPR Certificate]\label{def-vipr}
    A \emph{VIPR certificate} for a given $IP$ is an object $\cert$ that contains the following attributes:
    {\footnotesize
    \begin{itemize}
        \item $RTP(\cert)$: either $\mathtt{infeasible}$ 
        or an interval $[lb, ub]$ for some $lb \in \R \cup \{-\infty\}$ and $ub \in \R \cup \{\infty\}$.
        \item $SOL(\cert)$: a finite (possibly empty) set of points in $\R^n$.
        \item $DER(\cert)$: a (possibly empty) list of derived constraints $C_{m+1}, \dots, C_d$ (with $d \triangleq m$ if $DER(\cert)$ is empty) such that for all $m+1 \le k \le d$, we have $\A(C_k) \subseteq \{i \in [m+1, d]: \reason(C_i) = \asm \}$ and the following holds:
        \begin{itemize}
            \item If $\reason(C_k) \in \{\asm, \sol\}$, then $\data(C_k) = \NULL$.
            \item If $\reason(C_k) \in \{\lin, \rnd\}$, then $\data(C_k)$ is a point in $\R^d$. 
            \item If $\reason(C_k) = \uns$, then $\data(C_k) \in [d]^4$. 
        \end{itemize}
    \end{itemize}
     }%
\end{definition}

Let $C$ be a derived constraint in $DER(\cert)$ for a VIPR certificate $\cert$ and MILP $IP$. If $\reason(C) = \asm$, then $C$ is said to be an \emph{assumption}, meaning that $C$ is freely added to the MILP, and will be paired with another constraint and removed via the \emph{unsplit} ($\uns$) reasoning later on. If $\reason(C) \in \{\lin, \rnd\}$, then $C$ is derived as a linear combination of previous constraints in $DER(\cert)$, where the multipliers are defined from the associated data $\data(C)$. If $C$ has reasoning $\sol$, then $C$ is simply a ``free'' constraint which is the objective bound from the best feasible solution in $SOL(\cert)$. Lastly, if $C$ has reasoning $\uns$, then the data of $C$ has the form $(i_1, l_1, i_2, l_2)$, where $i_1, i_2$ indicate the constraints $C_{i_1}, C_{i_2}$ being unsplit, and $C_{l_1}, C_{l_2}$ are assumption constraints; the validity condition specified in Definition \ref{def-der-valid} further refines this intuition.

\begin{example}\label{example-cert}
    Let us present an example VIPR certificate for the running example MILP $IP_0$ from Example \ref{example-running-IP}. We will elaborate on the validity of this certificate after defining validity. 
    For illustrative purposes, instead of writing the data for a constraint with reasoning in $\{\lin, \rnd\}$, we instead write the linear combination that the data represents. So, the point $(2, 1, 0, -1)$ is written as $2 \times C_1 + C_2 + (-1) \times C_4$.
    Define the VIPR certificate $\cert_0$ for $IP_0$ with the following attributes.
    \begin{itemize}
        \item $RTP(\cert_0) = \infeasible$.
        \item $SOL(\cert_0) = \emptyset$.
        \item $DER(\cert_0)$ defined by the list of rows of Table \ref{tab:der-example}.
\begin{table}[h!]
    \centering
    \begin{tabular}{|c|c|c|c|c|}
        \hline
        Index & $C$ & $\reason(C)$ & $\data(C)$ & $\A(C)$ \\ \hline
        4 & $x_1 \le 0$ & $\asm$ & $\NULL$ & $\{4\}$ \\ \hline
        5 & $x_1 \ge 1$ & $\asm$ & $\NULL$ & $\{5\}$ \\ \hline
        6 & $x_2 \le 0$ & $\asm$ & $\NULL$ & $\{6\}$ \\ \hline
        7 & $0 \ge 1$ & $\lin$ & $C_1 + (-2) \times C_4 + (-3) \times C_6$ & $\{4, 6\}$ \\ \hline
        8 & $x_2 \ge 1$ & $\asm$ & $\NULL$ & $\{8\}$ \\ \hline
        9 & $0 \ge 1$ & $\lin$ & $(-\frac{1}{3}) \times C_3 + (-\frac{1}{3}) \times C_4 + 2 \times C_8$ & $\{4, 8\}$ \\ \hline
        10 & $x_2 \ge \frac{1}{4}$ & $\lin$ & $(-\frac{1}{4}) \times C_2 + (\frac{3}{4}) \times C_5$ & $\{5\}$ \\ \hline
        11 & $x_2 \ge 1$ & $\rnd$ & $C_{10}$ & $\{5\}$ \\ \hline
        12 & $0 \ge 1$ & $\lin$ & $(-\frac{1}{3}) \times C_2 + (-1) \times C_3 + \frac{14}{3} \times C_{11}$ & $\{5\}$ \\ \hline
        13 & $0 \ge 1$ & $\uns$ & $(7, 6, 9, 8)$ & $\{4\}$ \\ \hline
        14 & $0 \ge 1$ & $\uns$ & $(12, 5, 13, 4)$ & $\emptyset$ \\ \hline
    \end{tabular}
    \captionsetup{font={footnotesize}}
    \caption{$DER(\cert_0)$ for a VIPR certificate for $IP_0$.}
    \label{tab:der-example}
\end{table}
\end{itemize}
\hfill $\square$
\end{example}

With the definition of a VIPR certificate established, we now define what it means for a VIPR certificate to be \textit{valid}, which will imply Theorem \ref{thm-vipr-validity}'s desirable guarantees.
The following definition is very technical, but for intuition, it can be safely understood as defining how a derived constraint is \textit{implied} by previous parts of the certificate and assumptions, for each type of constraint implication reasoning.
For the following definition and in the next section, if $C$ is an ordinary (problem) constraint, we let $\A(C) = \emptyset$.

\begin{definition}[Valid Derived Constraint]\label{def-valid-der-constraint}
    Let $\cert$ be a VIPR certificate for a given $IP$. A derived constraint $C_{k}$ for $m+1 \le k \le d$ is \emph{valid} with respect to $(IP, \cert)$
    if the following conditions hold. 
    {\footnotesize
    \begin{itemize}
        \item If $\reason(C_k)= \asm$, then $\A(C_k) = \{k\}$.
        \item If $\reason(C_k)= \lin$, then $\data(C_k)$ satisfies $\nz (\data(C_k)) \subseteq [k-1]$ and generates a suitable linear combination of $C_{1}, \dots, C_d$ 
        such that the resulting constraint $C$ 
        dominates $C_k$, and $\A(C_k) = \bigcup_{i \in \nz(\data(C_k))} \A(C_i)$.
        \item If $\reason(C_k)= \rnd$, then $\data(C_k)$ satisfies $\nz (\data(C_k)) \subseteq [k-1]$ and generates a suitable linear combination of $C_1, \dots, C_d$ such that given the resulting constraint $C$, 
         we have that $\rnd(C)$ dominates $C_k$ 
         and $\A(C_k) = \bigcup_{i \in \nz(\data(C_k))} \A(C_i)$.
        \item If $\reason(C_k)= \uns$, then 
        $\data(C_k) = (i_1, l_1, i_2, l_2) \in [k-1]^4$ such that
        $C_{i_1}$ and $C_{i_2}$ both dominate $C_k$, 
        and the constraints $C_{l_1}, C_{l_2}$ form a split disjunction, and $\A(C_k) = (\A(C_{i_1}) \setminus \{l_1\}) \cup (\A(C_{i_2}) \setminus \{l_2\})$.
        \item If $\reason(C_k)= \sol$, then $\A(C_k) = \emptyset$ and there exists some $s \in SOL$ such that 
        \begin{itemize}
            \item if $\sense(IP) = \mathtt{min}$, then $c \cdot x \le c \cdot s$ dominates $C_k$, and
            \item if $\sense(IP) = \mathtt{max}$, then $c \cdot x \ge c \cdot s$ dominates $C_k$. 
        \end{itemize}
    \end{itemize}
    }%
\end{definition}

When $IP$ and $\cert$ are clear from context, for brevity we may simply say that a derived constraint $C_k$ is \emph{valid}, instead of valid with respect to $(IP, \cert)$.

The following definition of SOL-validity of a VIPR certificate simply asserts that the bound of the relation to prove (RTP) that can be proven by feasible solutions is achieved by at least one point in the feasible solution set $SOL$ (and that every solution in $SOL$ is indeed feasible). In the case of a minimization problem, upper bounds can be proven from feasible solutions, so we require that when the upper bound in RTP is $ub \ne \infty$, then some feasible solution $x$ in $SOL$ achieves this upper bound of $c \cdot x \le ub$. The other cases are defined similarly.



\begin{definition}[SOL-valid]\label{def-sol-valid}
    A VIPR certificate $\cert$ for a given $IP$ is \emph{SOL-valid} if $RTP = \infeasible$ and $SOL = \emptyset$, or if $RTP \ne \infeasible$ and all $x \in SOL$ are feasible in $IP$, and the following holds:
  {\footnotesize \begin{itemize}
        \item If $\sense(IP) = \mathtt{min}$ and $ub \ne \infty$, then at least one $x \in SOL$ has objective value $c \cdot x \le ub$.
        \item If $\sense(IP) = \mathtt{max}$ and $lb \ne -\infty$, then at least one $x \in SOL$ has objective value $c \cdot x \ge lb$.
    \end{itemize}
 }%
\end{definition}

We now define DER-validity below, which is designed to prove the dual bound to feasibility, or to prove infeasibility. This is done by requiring that all derived constraints are indeed implied by the problem constraints and its assumptions, and proving that the final derived constraint $C_d$ trivially implies the desired dual bound or infeasibility in the relation to prove. Specifically, $C_d$ should not depend on any ``assumption'' constraints, so that $C_d$ is implied by the problem constraints, and is shaped as the RTP requires.

\begin{definition}[DER-valid]\label{def-der-valid}
    A VIPR certificate $\cert$ for a given $IP$ is \emph{DER-valid} if the following holds.
   {\footnotesize \begin{itemize}
        \item Every derived constraint $C_i$, for $m+1 \le i \le d$, is valid.
        \item If $RTP = \infeasible$, then the last constraint $C_d$ (a \emph{derived constraint} unless $d = m$) is an absurdity with no assumptions (that is $\A(C_d) = \emptyset$).
        \item If $RTP \ne \infeasible$ and $\sense(IP) = \mathtt{min}$ and $lb \ne -\infty$, then $C_d$ dominates the inequality $c \cdot x \ge lb$ and $\A(C_d) = \emptyset$. 
        \item If $RTP \ne \infeasible$ and $\sense(IP) = \mathtt{max}$ and $ub \ne \infty$, then $C_d$ dominates the inequality $c \cdot x \le ub$ and $\A(C_d) = \emptyset$.
    \end{itemize}
    }%
\end{definition}

Finally, we are ready to define the validity conditions for a VIPR certificate, which simply enforces both SOL-validity and DER-validity. The consequences of VIPR certificate validity are given precisely in Theorem \ref{thm-vipr-validity}. Intuitively, validity implies that the optimality bounds or infeasibility given in the relation-to-prove are indeed correct.
\begin{definition}[Valid VIPR Certificate]\label{def-vipr-valid}
    A VIPR certificate $\cert$ for a given $IP$ is \emph{valid} if $\cert$ is SOL-valid and DER-valid.
\end{definition}

\begin{example}\label{example-cert-valid}
    Here, we show that the certificate $\cert_0$ in Example \ref{example-cert} for the MILP $IP_0$ in Example \ref{example-running-IP} is valid. It is easy to see that $\cert_0$ is SOL-valid since $RTP(\cert_0) = \infeasible$ and $SOL(\cert_0) = \emptyset$. To show that $\cert_0$ is DER-valid, it is easy to check that for every derived constraint $C_k \in DER(\cert_0)$, the conditions for derived constraint validity (Definition \ref{def-valid-der-constraint} are met on $C_k$; additionally, the last constraint in $DER(\cert_0)$ is the absurdity $0 \ge 1$ with no assumptions. Hence $\cert_0$ is DER-valid, and thus, valid. \hfill $\square$
\end{example}

Although in general the branch-and-cut framework for solving MILP instances is well-understood~\cite{ilp_theory}, for completeness we provide a self-contained argument for Theorem~\ref{thm-vipr-validity} given our context and notation.
\begin{theorem}\label{thm-vipr-validity}
    If a VIPR certificate $\cert$ for a given $IP$ is valid, then
    {\footnotesize
    \begin{itemize}
        \item if $RTP = \infeasible$, then $IP$ is infeasible, and
        \item if $RTP = [lb, ub]$, then the optimal objective value of $IP$ is in $[lb, ub]$ if an optimal 
        solution exists; if also $\sense(IP) = \mathtt{min}$ and $ub \neq \infty$, or if $\sense(IP) = \mathtt{max}$ and $lb \neq -\infty$, then $IP$ has a feasible solution with objective value in $[lb, ub]$.
    \end{itemize}
    }%
\end{theorem}
\begin{proof}
    See \ref{appen-vipr-thm}.
\end{proof}

Note the following consequence of Theorem \ref{thm-vipr-validity}. If $\sense(IP) = \mathtt{min}$ and $RTP = [-\infty, ub]$ for some $ub \in \R$, then $IP$ has a feasible solution with objective value at most $ub$, so the optimal objective value is at most $ub$ or $IP$ is unbounded (feasible but not optimal). 
A similar remark holds for when $\sense(IP) = \mathtt{max}$.

\subsection{Differences in Specifications and Abstraction}\label{subsec-differences}
The original \texttt{VIPR} certificate has a slightly different file format\footnote{\url{https://github.com/ambros-gleixner/VIPR/blob/master/cert_spec_v1_0.md}} than what we present here, but contains roughly the same information. The \texttt{VAR}, \texttt{INT}, \texttt{OBJ}, \texttt{CON} sections of the original specifications are all abstracted away in the definition of $IP$. Another difference is that no explicit information about assumptions of derived constraints is in the original \texttt{VIPR} format to save space; rather, the set of assumptions are implied and computed before executing the check. Furthermore, in the original specifications, the validity conditions for the set of assumptions in each derived constraint are described in informal language. We remove this possible ambiguity by treating the set of assumptions as precomputed and explicitly stated in Definition~\ref{def-vipr}, thus making precise how it is handled for each possible $\reason$ in Definition~\ref{def-valid-der-constraint}. Also in Definition~\ref{def-valid-der-constraint} we explicitly state general conditions for the validity of a derived constraint with $\sol$ reasoning, which are missing in the original specifications. 

The lack of an $\mathtt{index}$ attribute in a derived constraint is another notable difference in our abstraction. This field indicates the last derived constraint making use of the current one. As seen in Definition~\ref{def-der-valid}, we must check that every derived constraint in a VIPR certificate is valid, using Definition \ref{def-valid-der-constraint}. However, \texttt{viprchk} deletes constraints \emph{on the fly} by relying on their associated \texttt{index}. The potential performance gains from reduced memory consumption are problematic from the perspective of a skeptical checker, since once a constraint is deleted it cannot be checked anymore. In fact, its presence can be used to manipulate the workflow of \texttt{viprchk} and induce it to invalidate correct derivations as we illustrate in Subsection~\ref{sec:detectionForgedManipulated}. More generally this is a consequence of our weaker notion of validity in Definition~\ref{def-vipr-valid}, where we entirely disregard the $\texttt{index}$ attribute which \texttt{viprchk} currently depends on.

\subsection{Example Certificates}\label{subsec-examples}
Here we showcase example certificates mentioned in Section~\ref{sec:intro} that \texttt{viprchk} mishandles. The certificates exploit unexpected MILP solver behaviour. 

\begin{example}\label{example1}
    Define a MILP $IP_1$ with variables $x_1, x_2$ by 
    \[
    \max\{x_1 + x_2 : 0 \le x_1 \le 1, 0 \le x_2 \le 1, x_1 \in \Z, x_2 \in \Z\}.
    \]
    Define a VIPR certificate $\cert_1$ for $IP_1$ with $RTP(\cert_1) = [1,1] = \{1\}$ and $SOL(\cert_1) = \{(0,1)\}$ and $DER(\cert_1)$ being a list containing the derived constraint $x_1 + x_2 \le 1$ with no data, $\sol$ reasoning and no assumptions (see Definition \ref{def-derived-constraint}). It is easy to see that the optimal objective value of $IP_1$ is $2$, not the objective value of 1 reported by $RTP(\cert_1)$. By Theorem~\ref{thm-vipr-validity}, $\cert_1$ is not a valid VIPR certificate for $IP_1$. Yet when this certificate is run using \texttt{viprchk} (formatted as \texttt{forged1.vipr} in \ref{App-ForgedManipuated}), $\cert_1$ is reported as valid. \hfill $\square$
\end{example}

\begin{example}\label{example2}
    Define a MILP $IP_2$ with variables $x_1, x_2$ by
    \[
    \max \{x_1 + x_2 : 1/3 \le x_1 \le 1/2, 1/3 \le x_2 \le 1/2, x_1 \in \Z, x_2 \in \Z \}.
    \]
    Define a VIPR certificate $\cert_2$ for the MILP $IP_2$ by $RTP(\cert_2) = [-\infty, 0]$, $SOL(\cert_2) = \emptyset$ and $DER(\cert_2)$ being an empty list. 
    Since none of the constraints in $IP_2$ dominate the objective bound constraint $x_1 + x_2 \le 0$, it follows that $\cert_2$ is not a valid VIPR certificate for $IP_2$ by Definition \ref{def-der-valid}.
    However, when $\cert_2$ is run with \texttt{viprchk}, the certificate (formatted as \texttt{forged2.vipr} in \ref{App-ForgedManipuated}) is misinterpreted and \texttt{viprchk} ambiguously validates the result: \texttt{viprchk} prints \texttt{Successfully checked solution for feasibility} even though no solutions are reported in $SOL(\cert_2)$ and the problem is infeasible. \hfill $\square$
\end{example}

\subsection{Permissive Specifications}\label{sec:permissive}
Referring back to our point about permissive specifications in Section~\ref{sec:intro}, we make the following observations about derived constraint validity.
(i) A derived constraint $C_k$ with reasoning $\lin$ or $\rnd$ may take linear combinations of constraints from different branches of the branch-and-cut tree; in such a scenario, the set of constraints corresponding to the assumptions of $C_k$ contains a split disjunction, so that no solution can satisfy all assumptions simultaneously. These linear combinations are very unlikely to be used by most MILP solvers, but \texttt{VIPR} specifications allow for this. Similarly a derived constraint $C_k$ with reasoning $\asm$ need not belong to a corresponding split disjunction as in Definition~\ref{def-disjunction}, and can be \emph{any} constraint.
(ii) Another example of permissiveness is seen in the definition of validity for derived constraints $C_k$ with reasoning $\uns$. In particular, we do not require that $l_1 \in \A(C_{i_1})$ and $l_2 \in \A(C_{i_2})$; if either of these conditions fail, unsplitting is redundant. In fact, if we require those to be true, then a number of certificates we tested in Section \ref{sec:expts} would not be validated according to this stronger notion of validity.

\section{Predicate Transformation}\label{sec:smt}
In this section, we present a quantifier-free predicate in the theory of Linear Integer Real Arithmetic, which we show to be equivalent to the validity of a given VIPR certificate. Furthermore in Subsection~\ref{sec:why3}, we discuss a verification of our main equivalency theorem using \texttt{Why3}'s deductive logic framework \cite{Why3}. For the remainder of this section, let $IP$ be a MILP and let $\cert$ be a VIPR certificate for $IP$; we will use the notation presented in Definition \ref{def-vipr}. All proofs to the lemmas and the main theorem of this section may be found in \ref{app-proofs}.

First, we define some simple notation used throughout this section.

\begin{definition}[Constraint Notation]\label{def-constr-notation}
    For each constraint $C_i$ for $1 \le i \le d$, we write $l(C_i) = a_{i,1}x_1 + \cdots + a_{i, n} x_n$ and $r(C_i) = b_i$.
\end{definition}

\begin{definition}[Problem Type]\label{def-problem-type}
    We denote by $P$, called the \emph{problem type of $IP$}, the value of the predicate ``$\sense(IP) = \mathtt{min}$.''
\end{definition}

\begin{definition}[Result Type]\label{def-result-type}
    We denote by $R$, called the \emph{result type of $\cert$}, the value of the predicate ``$RTP \ne \infeasible$.''
\end{definition}


The following definition of $PUB$ (resp. $PLB$) captures when one must \textit{prove the upper bound} (resp. \textit{prove the lower bound}) of the RTP. Similarly, we define $U$ and $L$ to be equal to the upper and lower bounds in the RTP, when they exist (and are set to an arbitrary value otherwise, in this case 0).

\begin{definition}[Handling infinite upper and lower bounds]\label{def-handling-inf}
    Define the Boolean constants $PUB$ and $PLB$ as follows:
    {\footnotesize \[
    PUB \triangleq R \AND (ub \ne \infty)
    \quad \text{and} \quad 
    PLB \triangleq R \AND (lb \ne -\infty).
    \]
    }%
    Furthermore,
    define reals $U$ and $L$ as follows:
    {\footnotesize
    \[
    U \triangleq \begin{cases}
        ub, & \text{if } PUB\\
        0, & \text{if } \neg PUB
    \end{cases}
    \quad \text{and} \quad
    L \triangleq \begin{cases}
        lb, & \text{if } PLB\\
        0, & \text{if } \neg PLB.
    \end{cases}
    \]
    }%
\end{definition}

The rest of this section is structured as follows. For each component of the predicate, we give an intuitive description of its meaning and its equivalence to a meaningful component of validity, followed by a formal definition, followed by a formally stated lemma that states its significant to certificate validity. 

In some definitions---specifically, definitions around derived constraints---the associated lemmas are have hypotheses that \textit{depend on the validity of previous parts of the certificate.} This illustrates in important theoretical insight: that our predicates for derived constraint validity \textit{cannot} be isolated, but instead are required to have a certain inductive structure (looking backward). This sacrifice (not looking forward) allowed us to considerably save on the size of the assumption predicates for $\lin$ or $\rnd$ reasoning in Definition \ref{def-asm-predicate}.

Let us now begin defining the construction of each part of the end predicate $\phi$.

The following definition simply captures the statement that every solution in $SOL$ is feasible, in a slightly more concise way from the trivial construction.
For this definition, recall that $I$ denotes the set of all $j \in [n]$ such that $x_j$ is an integer variable.
\begin{definition}[Feasibility Predicate]\label{def-feas-predicate}
    For each $sol \in SOL$ and $i \in [m]$, define a predicate $\formula_{FEAS}(sol, i)$ as:
    {\footnotesize
    \begin{align*}
     & (s(C_i) \ge 0 \implies a_{i,1}sol_1 + \cdots + a_{i, n} sol_n \ge b_i)\\
     \AND  &(s(C_i) \le 0 \implies a_{i,1}sol_1 + \cdots + a_{i, n} sol_n \le b_i).
    \end{align*}
    }%
    Additionally, define $\formula_{FEAS}(sol) \triangleq \bigAnd_{j \in I} (sol_j \in \Z) \AND  \bigAnd_{i \in [m]} \formula_{FEAS}(sol,i)$. Finally, define
    $\formula_{FEAS} \triangleq \bigAnd_{sol \in SOL} \formula_{FEAS}(sol)$.
\end{definition}
\begin{lemma}\label{lem-feas}
    Every solution in $SOL$ is feasible if and only if $\formula_{FEAS} = \true$.
\end{lemma}

The following construction of the SOL-validity predicate simply captures the definition of SOL-validity in terms of the Boolean constants described in the beginning of this section.
\begin{definition}[SOL-valid Predicate]\label{def-sol-valid-predicate}
    Define $\formula_{SOL}$ as follows:
    {\footnotesize
    \begin{align*}
       & \ttif \neg R \ttthen |SOL| = 0 \ttelse \Biggl(\formula_{FEAS} \AND\\
       &\Biggl(\ttif P \ttthen \Biggl(PUB \implies \bigOr_{sol \in SOL} (c_1 sol_1 + \cdots + c_n sol_n \le U)\Biggr)\\
       & \ttelse \Biggl(PLB \implies \bigOr_{sol \in SOL} (c_1 sol_1 + \cdots + c_n sol_n \ge L)\Biggr) \Biggr) \Biggr).
    \end{align*}
    }%
\end{definition}
\begin{lemma}\label{lem-sol}
    The VIPR certificate $\cert$ is SOL-valid if and only if $\formula_{SOL} = \true$.
\end{lemma}

Now let us begin constructing a predicate that is true if and only if $\cert$ is DER-valid.

First, we define the assumption predicates below, which capture the correctness of the assumption sets for a derived constraint. This construction exploits the aforementioned observation that derived constraint validity predicates need not be correct in isolation, but need only be correct inductively.
Specifically, consider the case where determining membership in $\A(C_k)$ takes the longest na\"ively; this is when the reasoning is $\lin$ or $\rnd$, as the assumption constraints here are $\A(C_k) = \bigcup_{i \in \nz(\data(C_k))} \A(C_i)$. Under the assumption that all $i \in \nz(\data(C_k))$ are less than $k$ (which is enforced by a separate predicate, $\phi_{PRV}$ constructed later), and assuming an induction hypothesis, each such $i$ satisfies $\A(C_i) \subseteq [i-1] \subseteq [k-1]$. Thus it suffices to simply check that no $j > k$ is in $\A(C_k)$, and separately evaluate the obvious membership correctness checks for all $j < k$. Additionally, we make the following observation, conditioned on $\nz(\data(C_k)) \subseteq [k-1]$, which is enforced elsewhere: for an assumption index $j$, we have $j \in \bigcup_{i \in \nz(\data(C_k))} \A(C_i)$ if and only if $j \in \A(C_i)$ for some $i \in \nz(\data(C_k))$ satisfying $j \le i < k$. This significantly reduces the number of assumption membership evaluations.

\begin{definition}[Assumption Predicate]\label{def-asm-predicate}
    Let $S = \{i \in [m+1, d]: \reason(C_i) = \asm\}$.
    For all $k \in [d]$, let $S_{< k} = \{j \in S: j < k\}$ and $S_{> k} = \{j \in S: j > k\}$.
    For each $k \in [d]$ and $j \in [d]$, let $A_k^j$ be a Boolean constant defined to be the value of the predicate $j \in \A(C_k)$.
    For all $m+1 \le k \le d$, let $\formula_{ASM}(k)$ be the conjunction of $\bigAnd_{j \in S_{> k}} (\neg A_k^j)$ and
    {\footnotesize
    \[
    \begin{cases}
        A_k^k \AND \bigAnd_{j \in S_{< k}} \neg A_k^j, & \text{if } \reason(C_k) = \asm\\
        \bigAnd_{j \in S_{< k}} \biggl(A_k^j = \bigOr_{i \in \nz(\data(C_k)): j \le i < k} A_i^j\biggr), & \text{if } \reason(C_k) \in \{\lin,\rnd\}\\
        \bigAnd_{j \in S_{< k}} (A_k^j = (A_{i_1}^j \AND (j \ne l_1)) \OR (A_{i_2}^j \AND (j \ne l_2))), & \text{if } \reason(C_k) = \uns\\
        \bigAnd_{j \in S_{< k}} \neg A_k^j, & \text{if } \reason(C_k) = \sol,
    \end{cases}
    \]}
    where we write $\data(C_k) = (i_1, l_1, i_2, l_2)$ if $\reason(C_k) = \uns$. 
\end{definition}

Motivated by the previous paragraph, let us analyze the savings for the assumption requirements of $\lin$ and $\rnd$ constraint validity, comparing the na\"ive strategy to our definition in Definition \ref{def-asm-predicate}.
Precisely, this method yields the following savings on the number of assumption membership computations. Na\"ively, for $C_k$ with reasoning $\lin$ or $\rnd$, we compute $\sum_{j \in S} 1+|\nz(\data(C_k))| = |S| + |S| \cdot |\nz(\data(C_k))|$ assumption membership evaluations. Using the condensed formula given from the above observations, written in Definition \ref{def-asm-predicate}, we obtain
$
|S_{>k}| + \sum_{j \in S_{<k}} (1+|\nz(\data(C_k)) \cap [j, k)|) = |S| + |S_{<k}|\cdot \mathrm{avg}_{j \in S_{<k}} |\nz(\data(C_k)) \cap [j, k)|
$ evaluations of assumption membership, a considerable savings; here $\mathrm{avg}_{x \in \Lambda} f(x) \triangleq \frac{1}{|\Lambda|}\sum_{x \in \Lambda} f(x)$ denotes the average value of $f(x)$ for a real-valued function $f$ over a domain $\Lambda$.

For assumption derived constraints $C_k$, the only constraint on validity is that its assumptions only include itself, which is captured by $\phi_{ASM}(k)$. Thus we have the following construction.
\begin{definition}[Derived Constraint Formula: $\asm$]\label{def-asm-constr-predicate}
    For all $m+1 \le k \le d$ such that $\reason(C_k) = \asm$, define a predicate $\formula_{DER}(k) \triangleq \formula_{ASM}(k)$.
\end{definition}
\begin{lemma}\label{lem-asm-valid}
    Let $m+1 \le k \le d$ such that $\reason(C_k) = \asm$. Then $C_k$ is valid if and only if $\formula_{DER}(k) = \true$.
\end{lemma}


\begin{definition}[Non-zero $\data$ Validity Predicate]\label{def-prv-predicate}
    Consider any $m+1 \le k \le d$ such that $\reason(C_k) \in \{\lin, \rnd\}$.
    Let $\formula_{PRV}(k)$
    be the predicate defined by
    {\footnotesize \begin{align*}
    &\bigAnd_{j \in \nz(\data(C_k))} (j < k). 
    \end{align*}
    }%
\end{definition}

The following lemma is straightforward from Definition~\ref{def-prv-predicate}. 
\begin{lemma}\label{lem-prv}
    For all $m+1 \le k \le d$ such that $\reason(C_k) \in \{\lin, \rnd\}$, we have $\formula_{PRV}(k) = \true$ if and only if $\nz(\data(C_k)) \subseteq [k-1]$.
\end{lemma}

We have the following lemma, which shows that the inductive hypothesis on derived constraint validity, together with $\phi_{PRV}(k)$, imply the correctness of assumption sets for $\lin$ and $\rnd$ constraints.
\begin{lemma}\label{lem-asm-for-lin-rnd}
    Suppose $m+1 \le k \le d$ such that $\reason(C_k) \in \{\lin, \rnd\}$. Furthermore, assume that $\formula_{PRV}(k) = \true$ and that all previous derived constraints $C_{m+1}, \dots, C_{k-1}$ are valid. Then $\A(C_k) = \bigcup_{i \in \nz(\data(C_k))} \A(C_i)$ if and only if $\formula_{ASM}(k) = \true$.
\end{lemma}

The construction of a predicate for general constraint domination is defined below, and is very technical. We only define this fully expanded predicate because it applies to a more general type of constraint, where the sign may be undefined, as is the case with general linear combinations. For uses outside of $\lin$ and $\rnd$ reasoning, constraints are well-defined, so the below paragraph describes a convenient shorthand for regular constraints.
\begin{definition}[Domination Predicate]\label{def-domination-predicate}
    Consider vectors $a,a' \in \R^n$, reals $b, b' \in \R$, and Boolean values $eq, geq, leq, eq', geq', leq' \in \{\true, \false\}$. Then define the following \emph{domination} predicate $\formula_{DOM}(a,b,eq,geq,leq,a',b',eq',geq',leq')$:
    {\footnotesize
    \begin{align*}
        &\Biggl(\bigAnd_{j \in [n]} (a_j = 0) \AND (\ttif eq \ttthen b \ne 0 \ttelse \ttif geq \ttthen b > 0 \ttelse \ttif leq \ttthen b < 0\\
        & \ttelse \false ) \Biggr) \OR \Biggl(\bigAnd_{j \in [n]} (a_j = a'_j) \AND (\ttif eq' \ttthen (eq \AND (b = b')) \ttelse \ttif geq'\\
        &\ttthen (geq \AND (b \ge b')) \ttelse \ttif leq' \ttthen (leq \AND (b \le b')) \ttelse \false ) \Biggr).
    \end{align*}
    }
\end{definition}
When $C, C'$ are constraints with known left hand side ($l(C) = a \cdot x, l(C') = (a') \cdot x$), right hand side, and sign, we often write $\formula_{DOM}(C, C')$ as shorthand for $
\formula_{DOM}(a, r(C), s(C) = 0, s(C) \ge 0, s(C) \le 0, a', r(C'), s(C') = 0, s(C') \ge 0, s(C') \le 0).
$

The lemma below is very technical, but essentially states that our construction works as expected.
\begin{lemma}\label{lem-domination}
    Let $a,a' \in \R^n$, $b, b' \in \R$, and let $eq, geq, leq, eq', geq', leq'$ be Boolean constants. If there exists some (unique) signs $s,s' \in \{0,1,-1\}$ such that 
    $
    (s = 0) = eq, (s \ge 0) = geq, (s \le 0) = leq, (s' = 0) = eq', (s' \ge 0) = geq', (s' \le 0) = leq',
    $
    then the constraint $C$ defined by $l(C) = a \cdot x, s(C) = s, r(C) = b$ dominates the constraint $C'$ defined by $l(C') = (a') \cdot x, s(C') = s', r(C') = b'$ if and only if
    $
    \formula_{DOM}(a,b,eq,geq,leq,a',b',eq',geq',leq') = \true.$
    
    If $\neg (eq \OR geq \OR leq)$, then $\formula_{DOM}(a,b,eq,geq,leq,a',b',eq',geq',leq') = \false$.
\end{lemma}

Now we construct the predicate for $\lin$ derived constraints. Our construction is technical, but follows naturally from previous predicate constructions and the definition of validity in this case.

In the following definition the constants $eq, geq, leq, A, B$ are placeholders for their associated (longer) expressions. 
\begin{definition}[Derived Constraint Valid: $\lin$]\label{def-lin-constr-predicate}
    Consider any $m+1 \le k \le d$ such that $\reason(C_k) = \lin$.
    Define the following Boolean constants:
    {\footnotesize
    \[
    eq \triangleq \bigAnd_{i \in \nz(\data(C_k))} (\data(C_k)_i s(C_i) = 0), \qquad geq \triangleq \bigAnd_{i \in \nz(\data(C_k))} (\data(C_k)_i s(C_i) \ge 0),
    \]
    \[
    leq \triangleq \bigAnd_{i \in \nz(\data(C_k))} (\data(C_k)_i s(C_i) \le 0).
    \]}
    Write 
    {\footnotesize\[
    A \triangleq \left( \sum_{i \in \nz(\data(C_k))} \data(C_k)_i \cdot a_{i,j} \right)_{j \in [n]}, \qquad B \triangleq \sum_{i \in \nz(\data(C_k))} \data(C_k)_i \cdot b_i.
    \]}
    Finally, define the predicate $\formula_{DER}(k)$ by
    {\footnotesize\begin{align*}
    &\formula_{ASM}(k) \AND \formula_{PRV}(k)\\
    &\AND \formula_{DOM}(A, B, eq, geq, leq, (a_{k,j})_{j \in [n]}, b_k, s(C_k) = 0, s(C_k) \ge 0, s(C_k) \le 0).
    \end{align*}}
\end{definition}
\begin{lemma}\label{lem-lin-valid}
    If $m+1 \le k \le d$ such that $\reason(C_k) = \lin$ and all previous derived constraints are valid, then $C_k$ is valid if and only if $\formula_{DER}(k) = \true$.
\end{lemma}

The following construction captures when a constraint is roundable.
\begin{definition}[Roundable Predicate]\label{def-roundable-predicate}
    Let $a \in \R^n$, $b \in \R$, and let $eq$ be a Boolean constant. Define a predicate $\formula_{RND}(a,b,eq)$ by
    {\footnotesize
    \begin{align*}
        &\bigAnd_{j \in I} (a_{j} \in \Z) \AND \bigAnd_{i \notin I} (a_{i} = 0) \AND (\neg eq). 
    \end{align*}
    }%
\end{definition}
\begin{lemma}\label{lem-roundable}
    For a constraint $C$ with $l(C) = a \cdot x, r(C) = b$, we have $C$ is roundable if and only if $\formula_{RND}(a,b, s(C) = 0) = \true$.
\end{lemma}

We now construct $\phi_{DER}(k)$ for reasoning $\rnd$. The construction is very similar to the construction for $\lin$ constraints. However, here we exploit the requirement that $C_k$ cannot be an equality unless the underlying weighted linear combination is an absurdity, to slightly reduce the size of the predicate. While we could trivially adapt the formula from the $\lin$ reasoning case to the $\rnd$ reasoning case, we are able to reduce the overlapping casework slightly by replacing $\phi_{DOM}$ with a refined subformula optimized for this case.
\begin{definition}[Derived Constraint Valid: $\rnd$]\label{def-rnd-constr-predicate}
    Let $m+1 \le k \le d$ such that $\reason(C_k) = \rnd$. 
    Define $A \in \R^n$, $B \in \R$, and Boolean constants $eq, geq, leq$ as in Definition \ref{def-lin-constr-predicate}.
    Define the predicate $\formula_{DER}(k)$ by
    {\footnotesize
    \begin{align*}
    &\formula_{ASM}(k) \AND \formula_{PRV}(k) \AND \formula_{RND}\left( A, B, eq \right) \AND
    \Biggl( \Biggl(\bigAnd_{j \in [n]} (A_j = 0) \AND (\ttif geq \ttthen B > 0\\ & \ttelse \ttif leq \ttthen B < 0
    \ttelse \false ) \Biggr)
    \OR \Biggl(\bigAnd_{j \in [n]} (A_j = a_{k,j}) \AND (\ttif s(C_k) = 0 \ttthen\\ & \false \ttelse \ttif s(C_k) = 1
    \ttthen (geq \AND (\ceil{B} \ge b_k)) \ttelse (leq \AND (\floor{B} \le b_k))) \Biggr) \Biggr).
    \end{align*}
    }

\end{definition}
\begin{lemma}\label{lem-rnd-valid}
    For all $m+1 \le k \le d$ such that $\reason(C_k) = \rnd$ and all previous derived constraints are valid, we have that $C_k$ is valid if and only if $\formula_{DER}(k) = \true$.
\end{lemma}

The following construction captures when a pair of constraints is a split disjunction, which is a necessary property for validity of $\uns$ (unsplit) constraints.
\begin{definition}[Split Disjunction Predicate]\label{def-disjunction-predicate}
    Consider constraints $C_i, C_j$ for $1 \le i, j \le d$. Let $\formula_{DIS}(i, j)$ be the value of the predicate
    {\footnotesize\begin{align*}
        &\bigAnd_{k \in [n]} (a_{i,k} = a_{j,k}) \AND \bigAnd_{k \in I} (a_{i,k} \in \Z) \AND \bigAnd_{k \notin I} (a_{i,k} = 0) \AND (b_i \in \Z) \AND (b_j \in \Z)\\
        &\AND (s(C_i) \ne 0 \AND s(C_i) + s(C_j) = 0) \AND (\ttif s(C_i) = 1 \ttthen b_i = b_j + 1 \ttelse b_i = b_j-1).
    \end{align*}
    }%
\end{definition}
\begin{lemma}\label{lem-disjunction}
    For all $1\le i, j \le d$, we have that $C_i$ and $C_j$ form a split disjunction if and only if $\formula_{DIS}(i, j) = \true$.
\end{lemma}

The following construction of $\phi_{DER}(k)$ for reasoning $\uns$ follows naturally from the definition of validity for $\uns$ constraints, and the correctness of the above components (specifically, $\phi_{ASM}(k), \phi_{DOM},$ and $\phi_{DIS}$).
\begin{definition}[Derived Constraint Valid: $\uns$]\label{def-uns-constr-predicate}
    Suppose $m+1 \le k \le d$ with $\reason(C_k) = \uns$ and $\data(C_k) = (i_1, l_1, i_2, l_2)$. Define the predicate $\formula_{DER}(k)$ as 
    {\footnotesize
    \begin{align*}
        & \formula_{ASM}(k) \AND 
        (k > i_1) \AND (k > i_2) \AND (k > l_1) \AND (k > l_2)  \\
        & \AND 
        \formula_{DOM}(C_{i_1}, C_k) \AND \formula_{DOM}(C_{i_2}, C_k) \AND
        \formula_{DIS}(l_1, l_2).
    \end{align*}
    }%
\end{definition}
\begin{lemma}\label{lem-uns-valid}
    For all $m+1 \le k \le d$ with $\reason(C_k) = \uns$ and all previous derived constraints are valid, we have $C_k$ is valid if and only if $\formula_{DER}(k) = \true$.
\end{lemma}

The following construction of $\phi_{DER}(k)$ for derived constraint validity for $\sol$ reasoning is naturally constructed from the definition of validity, and the correctness of previously defined predicates.

\begin{definition}[Derived Constraint Valid: $\sol$]\label{def-sol-constr-predicate}
    Suppose $m+1 \le k \le d$ with $\reason(C_k) = \sol$. Define the predicate $\formula_{DER}(k)$ as
    {\footnotesize \begin{align*}
    \formula_{ASM}(k) \AND \Biggl(&\ttif P \ttthen \Biggl(\bigOr_{sol \in SOL} \formula_{DOM}(c \cdot x \le c \cdot sol, C_k)\Biggr)\\
    &\ttelse \Biggl(\bigOr_{sol \in SOL} \formula_{DOM}(c \cdot x \ge c \cdot sol, C_k)\Biggr)\Biggr).\\
    \end{align*}
    }%
\end{definition}
\begin{lemma}\label{lem-sol-constr-valid}
    For all $m+1 \le k \le d$ with $\reason(C_k) = \sol$, we have $C_k$ is valid if and only if $\formula_{DER}(k) = \true$.
\end{lemma}

The following corollary follows from Lemmas \ref{lem-asm-valid}, \ref{lem-lin-valid}, \ref{lem-rnd-valid}, \ref{lem-uns-valid}, \ref{lem-sol-constr-valid}.
\begin{corollary}\label{cor-der}
    For all $m+1 \le k \le d$ such that all previous derived constraints are valid, we have $C_k$ is valid if and only if $\formula_{DER}(k) = \true$.
\end{corollary}

We are finally ready to define a predicate to capture the DER-validity of a VIPR certificate. This predicate is defined below, and is constructed naturally from the definition of DER-validity. Its correctness, formalized in the lemma below, follows from the above corollary, and the equivalence of $\phi_{DOM}$ and constraint domination.
\begin{definition}[DER-valid Predicate]\label{def-der-predicate}
    Define the predicate $\formula_{DER}$
    {\footnotesize
    \begin{align*}
        &\left(\bigAnd_{m+1 \le k \le d} \formula_{DER}(k)\right)
        \AND \Biggl(\ttif \neg R \ttthen \left(\phi_{DOM}(C_d, 0 \ge 1) \AND \bigAnd_{j \in S} (\neg A_d^j)\right) \ttelse \\
        & \Biggl( \Biggl( (P \AND PLB) \implies \Biggl(\formula_{DOM}(C_d, c \cdot x \ge L)\AND \bigAnd_{j \in S} (\neg A_d^j) \Biggr) \Biggr)\\
        &\AND
        \Biggl( (\neg P \AND PUB) \implies \Biggl(\formula_{DOM}(C_d, c \cdot x \le U) \AND \bigAnd_{j \in S} (\neg A_d^j) \Biggr) \Biggr) \Biggr) \Biggr).
    \end{align*}
    }%
\end{definition}
\begin{lemma}\label{lem-der-valid}
    The VIPR certificate $\cert$ is DER-valid if and only if $\formula_{DER} = \true$. 
\end{lemma}

Finally, we arrive at our main construction and theorem.
\begin{theorem}[Equivalence]\label{thm-equivalence}
    The VIPR certificate $\cert$ is valid if and only if $\formula \triangleq \formula_{SOL} \AND \formula_{DER}$ 
    is true. 
\end{theorem}

\begin{example}\label{example-smt}
    Let us illustrate how the different parts of the VIPR certificate $\cert_0$ from Example \ref{example-cert} are translated into the predicate $\formula$ in Theorem \ref{thm-equivalence}.

    \begin{itemize}
        \item The clause $\formula_{SOL}$ that checks for SOL-validity will clearly evaluate to true, since $|SOL(\cert_0)| = 0$ and $\neg R = (RPT(\cert_0) = \infeasible) = \true$.
        
        \item The final $\ttif \ttthen \ttelse$ clause in $\formula_{DER}$ will evaluate to true, since the last derived constraint $0 \ge 1$ has no assumptions, clearly dominates $0 \ge 1$, and as above, $\neg R = \true$. See Table \ref{tab:der-example} for the list of the derived constraints in $\cert_0$.
        
        \item Lastly, for each derived constraint in $\cert_0$, Table \ref{tab:predicate-example} below points to the appropriate construction (definition number); the resulting predicates can be routinely checked by hand or with a solver, showing DER-validity, and thus, validity of $\cert_0$ with respect to the MILP $IP_0$.
    \end{itemize}

    \begin{table}[h!]
    \centering
    \begin{tabular}{|c|c|c|c|}
        \hline
        Index & $C$ & $\reason(C)$ & Construction \\ \hline
        4 & $x_1 \le 0$ & $\asm$ & Definition \ref{def-asm-constr-predicate} \\ \hline
        5 & $x_1 \ge 1$ & $\asm$ & Definition \ref{def-asm-constr-predicate} \\ \hline
        6 & $x_2 \le 0$ & $\asm$ & Definition \ref{def-asm-constr-predicate} \\ \hline
        7 & $0 \ge 1$ & $\lin$ & Definition \ref{def-lin-constr-predicate} \\ \hline
        8 & $x_2 \ge 1$ & $\asm$ & Definition \ref{def-asm-constr-predicate} \\ \hline
        9 & $0 \ge 1$ & $\lin$ & Definition \ref{def-lin-constr-predicate} \\ \hline
        10 & $x_2 \ge \frac{1}{4}$ & $\lin$ & Definition \ref{def-lin-constr-predicate} \\ \hline
        11 & $x_2 \ge 1$ & $\rnd$ & Definition \ref{def-rnd-constr-predicate} \\ \hline
        12 & $0 \ge 1$ & $\lin$ & Definition \ref{def-lin-constr-predicate} \\ \hline
        13 & $0 \ge 1$ & $\uns$ & Definition \ref{def-uns-constr-predicate} \\ \hline
        14 & $0 \ge 1$ & $\uns$ & Definition \ref{def-uns-constr-predicate} \\ \hline
    \end{tabular}
    \captionsetup{font={footnotesize}}
    \caption{Constructions for each $\formula_{DER}(k)$ for $\cert_0$.}
    \label{tab:predicate-example}
\end{table}
\end{example}

\subsection{Deductive Verification using \texttt{Why3}}\label{sec:why3}
\texttt{Why3} \cite{Why3} is a platform for deductive logic verification which provides the user with a rich language for writing logic specifications and programming, namely \texttt{WhyML}. In order to discharge verification conditions the platform relies on external theorem provers, both automated (SMT solvers) and interactive (Coq, Isabelle/HOL). The logic fragment of \texttt{Why3} is formalized in Coq~\cite{10.1145/3632902}, and the platform has been successfully used both in academic and industrial settings~\cite{chareton2021automated,da2020whylson}. In this work, in addition to traditional pen and paper arguments, we formalize the semantics of VIPR certificate validity in \texttt{Why3} and formally verify all core lemmas and equivalence theorem in Section~\ref{sec:smt} \cite{GitHub-Pulaj}. 
\begin{listing}[!ht]
\begin{minted}
[
frame=lines,
framesep=2mm,
baselinestretch=1.2,
bgcolor=light-gray,
fontsize=\footnotesize,
linenos=false,
fontsize=\scriptsize, breaklines,mathescape]{fstar}
type certificate = {
    cINT: set int;
    cOBJ: objective_function;
    cCON: array constraint;
    cRTP: rtp_type;
    cSOL: set (array real); 
    cDER: array derived_constraint;
    }
\end{minted}
\captionsetup{font={footnotesize}}
\caption{Definition of certificate type}

\label{listing:2}
\end{listing}

\newpage
\texttt{Why3} is based on first-order logic and allows for recursive definitions and algebraic data types which in our setting readily lend themselves to the formalization of VIPR certificates and related definitions (see for example Listing~\ref{listing:2}). In particular certificate attributes and validity semantics are captured in corresponding predicates (see for example Listing~\ref{listing:1}).
\begin{listing}[!ht]
\begin{minted}
[
frame=lines,
framesep=2mm,
baselinestretch=1.2,
bgcolor=light-gray,
fontsize=\footnotesize,
linenos=false,
fontsize=\scriptsize, breaklines,mathescape]{fstar}
  predicate is_cert (cert: certificate) =
    (forall i: int. mem i cert.cINT -> in_range i n) /\
    (is_objective_function cert.cOBJ) /\ (length cert.cCON = m) /\ 
    (forall i: int. in_range i m -> is_constraint cert.cCON[i]) /\
    (forall sol: (array real). mem sol cert.cSOL -> length sol = n) /\ 
    (length cert.cDER = (d - m)) /\ 
    (forall idx: int. in_range idx (length cert.cDER) -> is_der_con cert.cDER[idx]) /\
    (forall idx: int. in_range idx (length cert.cDER) -> 
      forall i: int. mem i cert.cDER[idx].a_set -> is_asm_index cert i)      
\end{minted}
\captionsetup{font={footnotesize}}
\caption{Predicate for attributes check in Definition ~\ref{def-vipr}}
\label{listing:3}
\end{listing}

\begin{listing}[hbt!]
\begin{minted}
[
frame=lines,
framesep=2mm,
baselinestretch=1.2,
bgcolor=light-gray,
fontsize=\footnotesize,
linenos=false,
fontsize=\scriptsize, breaklines,mathescape]{fstar}
predicate phi_DIS (cert: certificate) (ci cj: constraint) =
    (forall k: int. in_range k n -> ci.a[k] = cj.a[k]) /\
    (forall k: int. mem k cert.cINT -> is_integer ci.a[k]) /\
    (forall k: int. (in_range k n /\ not mem k cert.cINT) -> (ci.a[k] = 0.0)) /\
    (is_integer ci.b) /\ (is_integer cj.b) /\
    (ci.s <> Eq /\ cj.s <> Eq /\ ci.s <> cj.s) /\
    (if ci.s = Geq then ci.b = cj.b +. 1.0 else ci.b = cj.b -. 1.0)
\end{minted}
\captionsetup{font={footnotesize}}
\caption{Predicate for Definition~\ref{def-disjunction-predicate} }
\label{listing:1}
\end{listing}


\begin{table}[hbt!]
\centering
\scriptsize
\setlength{\tabcolsep}{3pt}
\renewcommand{\arraystretch}{0.95}
\begin{tabularx}{\linewidth}{l X}
\toprule
\textbf{Paper ref.} & \textbf{Why3 formalization} \\
\midrule
Def.~1–4 & \code{sign\_to\_int\_def}\sepa\code{is\_absurdity}\sepa \code{con\_dom\_con}\sepa \code{is\_disjunction} \\
Def.~5, 6, 8, 9 & \code{roundable},\code{rnd}\sepa \code{get\_lin\_comb}\sepa \code{derived\_constraint},\code{is\_der\_con}\sepa \code{is\_cert} \\
Def.~10–14 & \code{valid\_DER\_k}\sepa \code{valid\_SOL}\sepa \code{valid\_DER}\sepa \code{valid}\sepa \code{constraint} \\
Def.~15–17 & \code{p}\sepa \code{r}\sepa \code{pub}, \code{u}\sepa \code{plb}, \code{l} \\
Def.~18–21 & \code{phi\_FEAS}\sepa \code{phi\_SOL}\sepa \code{phi\_ASM} \\
Def.~23–27 & \code{phi\_DOM\_expanded} \sepa \code{phi\_DER\_k}(Lin) \sepa \code{phi\_RND}\sepa \code{phi\_DER\_k}(Rnd)\sepa \code{phi\_DIS} \\
Def.~30 & \code{phi\_DER} \\
\midrule
Lem.~1–3 & \code{LemmaFEAS} \sepa \code{LemmaSOL}\sepa\code{LemmaASM} \\
Lem.~4–5 & (subsumed) by \code{LemmaLin}, \code{LemmaRND} \\
Lem.~6–9 & \code{LemmaDOM}\sepa \code{LemmaLin}\sepa \code{LemmaRND\_Predicate}\sepa \code{LemmaRND} \\
Lem.~10–13 & \code{LemmaDIS}\sepa \code{LemmaUNS}\sepa \code{LemmaSOL\_valid}\sepa \code{LemmaDER} \\
Cor.~1 & \code{LemmaDER\_k} \\
Thm.~(Equiv.) & \code{MainTheorem} \\
\bottomrule
\end{tabularx}
\captionsetup{font={footnotesize}}
\caption{Overview of core formalization and corresponding pen and paper arguments }
\label{table:overview}
\end{table}
\vspace{0.5cm}

Other predicates capture the construction of the ground formula that is equivalent to the validity of a VIPR certificate (see for example Listing~\ref{listing:1}). The bulk of the formalization consists of the necessary type definitions and logical predicates, together with all the lemmas needed to prove equivalence. 
For a succinct overview of the correspondence between the formalization and pen and paper arguments we refer the reader to Table~\ref{table:overview}.

The proof effort of the formalization consists of about 130 lemmas, the majority of which are helper lemmas to guide the theorem provers. Many of the proofs are \emph{semi-automatic}: standard transformations in \texttt{Why3} (such as \texttt{unfold}, various versions of \texttt{split\_*}, \texttt{inline\_goal} or \texttt{induction}) need to be applied before the theorem provers can discharge the proof obligations. In particular, since the context becomes rather large, in lemmas leading to the equivalency theorem (see Listing~\ref{listing: 4}) clearing irrelevant context with the \texttt{clear\_but} transformation becomes crucially important in reducing runtime.

\begin{listing}[hbt!]
\begin{minted}
[
frame=lines,
framesep=2mm,
baselinestretch=1.2,
bgcolor=light-gray,
fontsize=\footnotesize,
linenos=false,
fontsize=\scriptsize, breaklines,mathescape]{fstar}
  lemma LemmaDER_k:
  forall cert: certificate, k: int.
    is_cert cert -> 0 <= k < d - m ->
    (forall j: int. 0 <= j < k -> valid_DER_k cert j) ->
    valid_DER_k cert k <-> phi_DER_k cert k 

  lemma LemmaDER: forall cert: certificate. is_cert cert ->
                    (valid_DER cert <-> phi_DER cert)  

  goal MainTheorem: forall cert: certificate. is_cert cert -> 
                      (valid cert <-> phi cert)                     
\end{minted}
\captionsetup{font={footnotesize}}
\caption{Corollary~\ref{cor-der}, Lemma ~\ref{lem-der-valid}, and Theorem~\ref{thm-equivalence}}
\label{listing: 4}
\end{listing}

In Table~\ref{tab:prover-stats} we provide summary runtime statistics for our formalization proof effort. For all additional details, including environment, set-up or reproducibility, we invite the reader to explore our GitHub repository\footnote{\href{https://github.com/JonadPulaj/vipr-gator/tree/main}{https://github.com/JonadPulaj/vipr-gator/tree/main}}.

\begin{table}[h]
\centering
\footnotesize
\setlength{\tabcolsep}{10pt} 
\begin{tabular}{lrrrrr}

\toprule
Prover & Att. & Succ. & Min & Max & Avg \\
\midrule
Alt-Ergo 2.5.4                & 250 & 250 & 0.01 & 21.27 & 1.10 \\
Alt-Ergo 2.5.4 (BV)           &  6 &   6 & 0.02 &  13.18 &  2.42 \\
CVC5 1.3.0                    & 787 & 787 & 0.02 & 23.44 & 0.31 \\
CVC5 1.3.0 (strings)          &  54 &  54 & 0.01 &  2.57 & 0.22 \\
Z3 4.15.2                     &   1 &   1 & 0.04 &  0.04 & 0.04 \\
Z3 4.15.2 (noBV)              &   10 &   10 & 0.10 &  7.58 & 1.32 \\
\bottomrule
\end{tabular}
\captionsetup{font={footnotesize}}
\caption{Prover statistics (attempts, successes, and times in seconds).}
\label{tab:prover-stats}
\end{table}
\subsection{Towards a safer checker for VIPR certificates}\label{sec:further_integ}
Formalizing the semantics of VIPR validity and proving its equivalence to a ground formula clearly paves the way for a safer checker and enables a significant step towards this goal. In this work we implement a complete parser (see Section~\ref{sec:expts} for technical details) which translates a VIPR certificate file to the formalized equivalency ground formula and \emph{validates} it with an external SMT-solver, although an SMT-solver is not strictly necessary for this as a model-checking engine such as Dolmen~\cite{bury2023verifying} suffices. 
There are no uninterpreted symbols or variables, so the SMT solver acts purely as a Boolean function evaluator in our setting. Because of this structure, unsatisfiable outcomes in the SMT-solver carry no additional semantic weight in our tool\footnote{ Usually UNSAT results are precisely the ones that need further verification, hence the need for SMTCoq~\cite{ekici2017smtcoq} for example. This is not the case in our setting, where if a certificate is malformed or incorrectly derived the failure is readily accessible.}---they simply lead to rejected certificates. 

Note that it is straightforward to check that any correctly rejected certificate is indeed \emph{not} valid as any such certificate will have at least one part of the conjunctive formula evaluating to false, which the runtime SMT solver catches. One can then evaluate the proposed false sub-formula by inspecting the certificate and verifying that the corresponding part of the predicate indeed evaluates to false.

As an alternative, one could completely bypass the use of SMT-solvers (or model-checking engines) and instead make direct use of \texttt{SMT Lean}~\cite{githubGitHubUfmgsmiteleansmt}, a project which enables Lean tactics to discharge goals into SMT solvers. The main available tactic transforms goals into SMT queries then delegates them to \texttt{cvc5}. Our implementation in Section~\ref{sec:expts} can be appropriately modified to write the formalized equivalency predicate directly into a \texttt{Lean} file, then use \texttt{SMT Lean} to discharge the proof obligations.

Finally a natural direction and extension of our current formalization work is mirroring the logical predicates in \texttt{Why3} with executable (non-ghost) functions that can be extracted (for example to \texttt{OCaml} code). Bridge lemmas can ensure the equivalence between the logical and executable code. This extension is a promising direction for future work.

\section{Experiments}

\label{sec:expts}

In this section, we evaluate an implementation of our analytical framework as discussed above, compared with $\texttt{viprchk}$. Similar to~\cite{vipr1}, our implementation is in C++, although we use C++20 extensions to allow much of the translation to be expressed in a functional style of programming. 
Our implementation consists of about 4,500 lines, and we run both our tool and $\texttt{viprchk}$ in a RHELinux 8.10, compiled with \texttt{clang} 17 with $-O3$ optimization turned on. We also provide the option of completely disregarding any parallelism, significantly reducing the trusted codebase. In the next subsection we discuss our general implementation strategies and optimizations, followed by comparative performance results and a section that discusses how our tool can detect some forged certificates which $\texttt{viprchk}$ currently cannot.

\subsection{Implementation Strategies and Optimizations}

After we tokenize and parse the input file, we maintain results in a \texttt{Certificate} object in memory. The parser uses a linear memory allocator that obtains 1GB memory chunks from the OS at a time, maintaining all strings that were contiguous in the input also contiguous in the memory, improving cache locality. 
Our C++ translation tool never checks statements, and only computes assumptions $A_i^j$ and  $\formula_{ASM}$ as well as $geq$, $eq$, and $leq$ according to their definitions. We do not generate multiplication with terms that have been parsed as zero. 
All other predicate evaluations are delegated to the SMT solver \texttt{cvc5}.

Now observe that our final formula is simply a large conjunction of Boolean expressions. A single machine generates the formula, but it leverages parallelism by using 64 cores to generate different parts of the formula. We generate one file for $\formula_{SOL}$, and multiple files that partition the $\bigAnd_{m+1 \le k \le d} \formula_{DER}(k)$ part of $\formula_{DER}$, each containing a subset of $D = \{\formula_{DER}(i): m+1 \le i \le d\}$, and one extra file for the final $(\ttif \ttthen \ttelse)$ expression in $\formula_{DER}$ (Definition \ref{def-der-predicate}). Specifically, the set of formulas in $D$ is divided into sets of size $\mathrm{blockSize}$ elements called \emph{blocks}, where $\mathrm{blockSize} \triangleq \floor{|D|/\mathrm{localCPUs}}$, with a potential extra set containing the remainder elements of $D$ if the division is not exact ($\mathrm{localCPUs}$ denotes the number of cores available locally for the file generation).

Once all files are generated, a fourth and final phase dispatches each of the files to all the hardware threads available in all machines in the network to be checked using \texttt{cvc5}. Now, not only the original $\mathrm{localCPUs}$ are available for computation, but a larger number of cores distributed in 3 machines ($\mathrm{totalCPUs}$ denotes the number of cores available in the whole network). The original program that generated the formulas waits for all answers and yields a final output of $\mathtt{sat}$ or $\mathtt{unsat}$. Our program is ready to exit cleanly if an $\mathtt{unsat}$ response is obtained from any of the remote dispatches. The correctness of this approach should be clear, since the predicate $\formula$ is precisely equal to the conjunction of $\formula_{SOL}$, all blocks, and the final $(\ttif \ttthen \ttelse)$ expression in Definition \ref{def-der-predicate}.



\subsection{Performance Results}

Our main experiment consists of all 106 \texttt{VIPR} instances\footnote{\url{https://github.com/ambros-gleixner/VIPR/tree/master/experiments}} from \cite{vipr1}. Table~\ref{tbl:results} presents the shifted geometric mean (by 10s) of the times indicated by the columns (as standard in the MILP community). ``Parse'' refers to the time to parse the certificate; ``Generations'' refers to the time to transform the certificate into an SMT instance; ``Verification'' refers to the time to verify the associated SMT instance; ``Total$_{SMT}$'' is the total time using our method; and ``Total$_{\texttt{viprchk}}$'' is the total time using {\texttt{viprchk}}. The problems are divided between 56 numerically easy instances and 50 numerically hard ones, although this naming convention does not correlate with their implied result, both in our tool and in \texttt{viprchk}. All of our checks confirm the validity of the \texttt{VIPR} certificates as in \cite{vipr1}. The full dataset is in \cite{GitHub-Mendes}, showing that we benefit from problems with a large number of derivations (as we generate/check them in parallel), but on the other hand files with a large number of constraints force us to produce much bigger SMT files in the disk. In any case, on average we perform checks in less time than what is reported to solve and produce certificates by \texttt{exact SCIP} (comparing our times to the reported \texttt{exact SCIP} times in \cite{vipr1}). In addition, our extra checks allow us to detect forged certificates that otherwise pass the \texttt{viprchk} tool.

\begin{table}[htb]
  \centering
  \caption{\label{fig3:t1} Breakdown of timing among different phases of our strategy and comparison with \texttt{viprchk}.}
\begin{tabular}{|c|c|c|c|c|c|}
\hline
Instances & Parse & Generation & Verification & Total$_{\text{SMT}}$ & Total$_{\texttt{viprchk}}$ \\ \hline
easy \hspace{1cm} & 5.772 & 21.494 & 63.079 & 104.060 & 42.152 \\ \hline
hard \hspace{1cm} & 14.857 & 36.887 & 46.692 & 82.513 & 19.173 \\ \hline
\end{tabular}
\label{tbl:results}
\end{table}

\subsection{Detection of Forged or Manipulated Certificates}
\label{sec:detectionForgedManipulated}

The certificates used to measure performance were all valid, but it is of course appropriate to test our implementation with certificates that are invalid. First, we instantiate the two invalid certificates from Section \ref{subsec-examples} that, as of July 2024, do not pass the \texttt{viprchk} tool, while our tool correctly identifies the equivalent formulas as $\mathtt{unsat}$. These are available in \cite{GitHub-Mendes} as well as in \ref{App-ForgedManipuated} with names \texttt{forged1.vipr} and \texttt{forged2.vipr}.
In addition, we provide an example of \texttt{viprchk}'s over-reliance on the \texttt{index} attribute of the derivations in an example called \texttt{manipulated1.vipr} (\cite{GitHub-Mendes} and \ref{App-ForgedManipuated}). This file changes the index in derivation \texttt{C8} of \texttt{paper\_eg3.vipr} in \cite{vipr1}, and \texttt{viprchk} fails to validate an instance that still contains only correct derivations and conclusions. Our tool correctly identifies that all derivations are sound. In addition, we introduced random changes to numbers in the certificate input of the files we tested, and we correctly identified that the associated derivations were unsound.

\section{Conclusion}
\label{sec:concl}

In this work we seek to leverage advances in SMT solvers to verify VIPR certificates for MILP solvers. Thus we design a VIPR certificate checker that transforms the logic of the VIPR certificate into an equivalent ground formula in the theory of Linear Integer Real Arithmetic. Furthermore we formally verify that the logic of the certificate is equivalent to a ground formula using \texttt{Why3}. We test the viability of our checker on benchmark instances in the literature. Looking forward, we seek to mirror non-ghost code from the logical code in the current \texttt{Why3} formalization to enable an extraction of a safe checker for VIPR certificates.


\section*{Acknowledgments}
The authors are very grateful for the careful and valuable feedback from the anonymous reviewers. Their insightful comments helped improve the exposition and, in particular, helped strengthen the current results by prompting the present formalization of VIPR certificates in \texttt{Why3} given in this work.

\section*{CRediT authorship contribution statement}
\textbf{Kenan Wood:} Conceptualization, Formal analysis, Investigation, Software, Writing -- original draft, Writing -- review and editing. \textbf{Runtian Zhou:} Conceptualization, Formal analysis, Investigation, Software, Writing -- original draft, Writing -- review and editing. \textbf{Haoze Wu:} Conceptualization, Investigation, Software, Writing -- review and editing. \textbf{Hammurabi Mendes:} Formal analysis, Software, Writing -- original draft, Writing -- review and editing. \textbf{Jonad Pulaj:} Conceptualization, Formal analysis, Investigation, Methodology, Software, Writing -- original draft, Writing -- review and editing.

\section*{Declaration of competing interest}
The authors declare that they have no competing interests that could have inappropriately bias or influence the work in this paper.

\bibliographystyle{plain}
\bibliography{main}
\newpage
\appendix
\section{Proof of Theorem \ref{thm-vipr-validity}}\label{appen-vipr-thm}
This section contains the proof of the main theorem in Section \ref{sec:vipr}, Theorem \ref{thm-vipr-validity}.



\begin{proof}[Proof of Theorem \ref{thm-vipr-validity}]
    Suppose $\cert$ is valid for $IP$.
    It suffices to prove the following claim, by the Definitions of SOL-validity and DER-validity: If $SOL = \emptyset$, then for all $i \in [d]$, every feasible solution $x \in \R^n$ that satisfies all constraints $C_j$ for $j \in \A(C_i)$ also satisfies $C_i$; if $SOL \ne \emptyset$, then this holds for all \emph{optimal} solutions $x$. 

    First, suppose $SOL = \emptyset$. Since every derived constraint is valid, no derived constraint has reasoning $\sol$. Let us proceed by induction on $k$. If $k \in [m]$, then the claim holds for $i = k$ by definition of feasibility. Suppose $m+1 \le k \le d$ and the claim holds for all $i \in [k-1]$. Then $\reason(C_k) \ne \sol$ by the observation above. Suppose $x \in \R^n$ is feasible and satisfies the constraints $C_j$ for $j \in \A(C_k)$. We then have the following cases.
    \begin{itemize}
        \item Suppose $\reason(C_k) = \asm$. Then $\A(C_k) = \{k\}$, so that $x$ satisfies $C_k$.
        
        \item Suppose $\reason(C_k) \in \{\lin, \rnd\}$. Then $\nz(\data(C_k)) \subseteq [k-1]$ and $\A(C_k) = \bigcup_{i \in \nz(\data(C_k))} \A(C_i)$. This implies that for all $i \in \nz(\data(C_k))$, we have $\A(C_i) \subseteq \A(C_k)$, so that $x$ satisfies $C_j$ for all $j \in \A(C_i)$. By the induction hypothesis, $x$ satisfies $C_i$ for all $i \in \nz(\data(C_k))$, which shows that $x$ satisfies $C \triangleq \sum_{i \in \nz(\data(C_k))} \data(C_k) \cdot C_i$ and also $\rnd(C)$ if $\reason(C_k) = \rnd$. If $\reason(C_k) = \lin$, then since $C$ dominates $C_k$, this shows that $x$ satisfies $C_k$. If $\reason(C_k) = \rnd$, then since $\rnd(C)$ dominates $C_k$, we know $x$ satisfies $C_k$.
        
        \item Suppose $\reason(C_k) = \uns$. Write $\data(C_k) = (i_1, l_1, i_2, l_2)$. Then $i_1, l_2, i_2, l_2$ are all less than $k$,
        and $\A(C_k) = (\A(C_{i_1}) \setminus \{l_1\}) \cup (\A(C_{i_2}) \setminus \{l_2\})$. Since $C_{l_1}$ and $C_{l_2}$ form a disjunction, we know that $x$ satisfies $C_{l_1}$ or $C_{l_2}$. If $x$ satisfies $C_{l_1}$, then since $\A(C_{i_1}) \setminus \{l_1\} \subseteq \A(C_k)$, we know that $x$ satisfies $C_i$ for all $i \in \A(C_{i_1})$; this implies that $x$ satisfies $C_{i_1}$ by the induction hypothesis, so that $x$ satisfies $C_k$ since $C_{i_1}$ dominates $C_k$. By analogous reasoning, if $x$ instead satisfies $C_{l_2}$, then still, $x$ satisfies $C_k$.
    \end{itemize}

    Now suppose that $SOL \ne \emptyset$. The proof of this case is almost identical, except noting that if some derived constraint $C_i$ has $\reason(C_i) = \sol$, then every \emph{optimal} solution $x$ satisfies $C_i$ since all solutions in $SOL$ are feasible (by SOL-validity).

    Now let us show that this claim implies the theorem. Suppose $RTP = \infeasible$. Then certainly $SOL = \emptyset$ (by definition of SOL-validity), so that every feasible solution satisfies $C_d$ (since $C_d$ has no assumptions); since $C_d$ is an absurdity, no feasible solutions exist, and so $IP$ is infeasible.

    On the other hand, suppose $RTP = [lb, ub]$. Then all optimal solutions satisfy $C_d$ since $C_d$ has no assumptions. First, suppose $\sense(IP) = \mathtt{min}$. If an optimal solution $x^*$ with objective value $v^*$ exists, $x^*$ is certainly feasible; by SOL-validity of $\cert$, $IP$ has a feasible solution with objective value at most $ub$ (this holds regardless of if $ub \in \R$ or $ub = \infty$). Hence $v^* \le ub$. If $lb = -\infty$, then we certainly have $v^* \ge lb$; otherwise, $C_d$ dominates the inequality $c \cdot x \ge lb$, so that $v^* = c\cdot x^* \ge lb$ as $x^*$ satisfies $C_d$. Hence $v^* \in [lb, ub]$. For the second part where $ub \ne \infty$, SOL-validity shows that $IP$ has a feasible solution with objective value at most $ub$; this objective value must be at least $v^* \ge lb$ by the minimality of $v^*$. The theorem thus follows when $\sense(IP) = \mathtt{min}$.

    The case when $\sense(IP) = \mathtt{min}$ follows from a similar argument.
\end{proof}
\section{Proof of SMT Transformation Correctness}\label{app-proofs}
Here, we write the proofs of each of the lemmas and the main theorem in Section \ref{sec:smt}.

\begin{proof}
    [Proof of Lemma \ref{lem-feas}]
    First, observe that for all $sol \in SOL$ and $i \in [m]$, we have the following: $sol$ satisfies constraint $C_i$ if and only if $\formula_{FEAS}(sol, i) = \true$ (this follows from case analysis on $s(C_i)$). It follows that $sol$ is feasible if and only if $\formula_{FEAS}(sol) = \true$. The lemma follows immediately.
\end{proof}

\begin{proof}[Proof of Lemma \ref{lem-sol}]
    This follows immediately from Lemma \ref{lem-feas} and Definition \ref{def-sol-valid}.
\end{proof}

\begin{proof}[Proof of Lemma \ref{lem-asm-valid}]
    By Definition \ref{def-vipr}, $\A(C_k) \subseteq S$. Hence $C_k$ is valid if and only if $\A(C_k) = \{k\}$, which holds if and only if $\formula_{DER}(k) = \true$, by Definitions \ref{def-asm-predicate} and \ref{def-asm-constr-predicate}.
\end{proof}

\begin{proof}[Proof of Lemma \ref{lem-asm-for-lin-rnd}]
    Immediately, we have $\A(C_k) \subseteq S$. Also, since $\formula_{PRV}(k) = \true$, Lemma \ref{lem-prv} shows that $\nz(\data(C_k)) \subseteq [k-1]$. 
    
    Suppose $\A(C_k) = \bigcup_{i \in \nz(\data(C_k))} \A(C_i)$. Since all constraints $C_{m+1}, \dots, C_{k-1}$ are valid, a simple induction argument shows that each $i \in \nz(\data(C_k))$ satisfies $\A(C_i) \subseteq [i]$, so $\A(C_k) \subseteq [k-1]$ as $\nz(\data(C_k)) \subseteq [k-1]$. Hence we have $\bigAnd_{j \in S_{>k}}(\neg A_k^j) = \true$. Additionally, for all $j \in S_{< k} \subseteq [k-1]$, we have $j \in \A(C_k)$ if and only if there exists $i \in \nz(\data(C_k))$ such that $j \in \A(C_i)$; since each $i \in \nz(\data(C_k))$ satisfies $\A(C_i) \subseteq [i]$, any $i\in \nz(\data(C_k))$ that satisfies $j \in \A(C_i)$ also satisfies $j \le i < k$. It follows that for all $j \in S_{< k}$, we have $A_k^j = \bigOr_{i \in \nz(\data(C_k)): j \le i < k} A_i^j$. Hence $\formula_{ASM}(k) = \true$.

    Conversely, suppose $\formula_{ASM}(k) = \true$. Then since $\bigAnd_{j \in S_{>k}} (\neg A_k^j) = \true$ and $\A(C_k) \subseteq S$, we have $\A(C_k) \subseteq S_{< k}$. Suppose $j \in \A(C_k)$, so $A_k^j = \true$ and $j \in S_{< k}$. Since $\formula_{ASM}(k) = \true$, we have $\bigOr_{i \in \nz(\data(C_k)): j \le i < k} A_i^j = A_k^j = \true$. In particular, this implies some $i \in \nz(\data(C_k))$ satisfies $j \in \A(C_i)$. It follows that $\A(C_k) \subseteq \bigcup_{i \in \nz(\data(C_k))} \A(C_i)$. Conversely, suppose $j \in \bigcup_{i \in \nz(\data(C_k))} \A(C_i)$. Immediately, $j \in S$. Also there exists some $i \in \nz(\data(C_k))$ such that $j \in \A(C_i)$. As $i < k$, we have $\A(C_i) \subseteq [i]$, so $j \le i < k$. It follows that $j \in S_{< k}$, so 
    \[
    A_k^j = \bigOr_{i \in \nz(\data(C_k)): j \le i < k} A_i^j = \true.
    \]
    Hence $j \in \A(C_k)$, as desired. Hence $\A(C_k) = \bigcup_{i \in \nz(\data(C_k))} \A(C_i)$.
\end{proof}

\begin{proof}[Proof of Lemma \ref{lem-domination}]
    The first part of this lemma follows immediately from Definition \ref{def-domination}, after observing that $geq \AND (\neg eq)$ is equivalent to $(s = 1)$ and $leq \AND (\neg eq) \AND (\neg geq)$ is equivalent to $s = -1$. The second part follows immediately by the construction in Definition \ref{def-domination-predicate}.
\end{proof}

\begin{proof}[Proof of Lemma \ref{lem-lin-valid}]
    Suppose all constraints $C_{m+1}, \dots, C_{k-1}$ are valid. Now suppose $C_k$ is valid. By Lemma \ref{lem-prv}, $\formula_{PRV}(k) = \true$. By Lemma \ref{lem-asm-for-lin-rnd}, since $\A(C_k) = \bigcup_{i \in \nz(\data(C_k))} \A(C_i)$, we have that $\formula_{ASM}(k) = \true$. Since $\data(C_k)$ generates a suitable linear combination of $C_1, \dots, C_d$ we have that $geq \OR leq = \true$ and that the linear combination $\sum_{i \in \nz(\data(C_k))} \data(C_k)_i C_i$ is well-defined. The construction implies that $eq = (s(\sum_{i \in \nz(\data(C_k))} \data(C_k)_i C_i) = 0)$ and $geq = (s(\sum_{i \in \nz(\data(C_k))} \data(C_k)_i C_i) \ge 0)$ and $leq = (s(\sum_{i \in \nz(\data(C_k))} \data(C_k)_i C_i) \le 0)$. Then since $\sum_{i \in \nz(\data(C_k))} \data(C_k)_i C_i$ dominates $C_k$, we have
    {\footnotesize\[
    \formula_{DOM}(A, B, eq, geq, leq, (a_{k,j})_{j \in [n]}, b_k, s(C_k) = 0, s(C_k) \ge 0, s(C_k) \le 0) = \true
    \]}
    by Lemma \ref{lem-domination}.
    It follows that $\formula_{DER}(k) = \true$.

    Conversely, suppose $\formula_{DER}(k) = \true$. Since $\formula_{PRV}(k) = \true$, Lemma \ref{lem-prv} implies that $\nz(\data(C_k)) \subseteq [k-1]$. Since also $\formula_{ASM}(k) = \true$, Lemma \ref{lem-asm-for-lin-rnd} shows that $\A(C_k) = \bigcup_{i \in \nz(\data(C_k))} \A(C_i)$. Now, as the domination predicate in $\formula_{DER}(k)$ is true, Lemma \ref{lem-domination} implies that at least one of $eq, geq, leq$ is true. As $eq \implies geq$ and $eq \implies leq$, at least one of $geq, leq$ is true, which shows that $\data(C_k)$ generates a suitable linear combination of $C_1, \dots, C_d$. Let $s = s(\sum_{i \in \nz(\data(C_k))} \data(C_k)_i C_i)$. By construction, $eq = (s = 0)$ and $geq = (s \ge 0)$ and $leq = (s \le 0)$. By Lemma \ref{lem-domination}, this shows that $\sum_{i \in \nz(\data(C_k))} \data(C_k)_i C_i$ dominates $C_k$ since $\formula_{DOM}(A, B, eq, geq, leq, (a_{k,j})_{j \in [n]}, b_k, s(C_k) = 0, s(C_k) \ge 0, s(C_k) \le 0) = \true$. It follows that $C_k$ is a valid derived constraint.
\end{proof}

\begin{proof}[Proof of Lemma \ref{lem-roundable}]
    This follows directly from Definition \ref{def-rounding} and the construction of $\formula_{RND}(a,b,eq)$.
\end{proof}

\begin{proof}[Proof of Lemma \ref{lem-rnd-valid}]
    Suppose $\reason(C_k) = \rnd$ and all derived constraints $C_{m+1}, \dots, C_{k-1}$ are valid. First, suppose $C_k$ is valid. We first verify that $\formula_{ASM}(k) = \true$, $\formula_{PRV}(k) = \true$, and $\formula_{RND}(A, B, eq) = \true$. Since $\nz(\data(C_k)) \subseteq [k-1]$, Lemma \ref{lem-prv} shows that $\formula_{PRV}(k) = \true$. By Lemma \ref{lem-asm-for-lin-rnd}, we have  $\formula_{ASM}(k) = \true$ since $\A(C_k) = \bigcup_{i \in \nz(\data(C_k))} \A(C_i)$. We also have that $\data(C_k)$ generates a suitable linear combination of $C_1, \dots, C_d$ such that $C \triangleq \sum_{i \in \nz(\data(C_k))} \data(C_k)_i C_i$ is roundable. By Lemma \ref{lem-roundable}, $\formula_{RND}(A,B,eq) = \true$.

    Now we show that the last conjunctive subpredicate of $\formula_{DER}(k)$ evaluates to $\true$. Since $C_k$ is valid, $\rnd(C)$ dominates $C_k$, where $C$ is defined as the linear combination above. We have two cases.
    \begin{itemize}
        \item Suppose $\rnd(C)$ is an absurdity. Clearly $\bigAnd_{j \in [n]} (A_j = 0) = \true$. It is easy to see that $C$ is also an absurdity. Since $\formula_{RND}(A,B,eq) = \true$, we have $eq = \false$, so that $s(C) = 1$, in which case $geq = \true$ and $leq = \false$, or $s(C) = -1$, in which case $leq = \true$ and $geq = \false$. Since $C$ is an absurdity, it follows that $(\ttif geq \ttthen B > 0 \ttelse \ttif leq \ttthen B < 0 \ttelse \false) = \true$, as desired. The claim follows.

        \item Suppose $\rnd(C)$ is not an absurdity. Then, by the definition of domination, $l(C) = l(\rnd(C)) = l(C_k)$ and $s(C) = s(\rnd(C)) = s(C_k)$. Hence $\bigAnd_{j \in [n]} (A_j = a_{k,j}) = \true$. Since $\formula_{RND}(k) = \true$, so $eq = \false$, we have $s(C) \ne 0$, so $s(C_k) \ne 0$. If $s(C_k) = 1$, then we must have $geq = \true$; by the definition of the rounding of a roundable constraint, namely $C$, we have $\ceil{B} \ge b_k$. Analogously, if $s(C_k) = -1$, then $leq \AND (\floor{B} \le b_k)$. The claim follows.
    \end{itemize}

    Conversely, let us suppose $\formula_{DER}(k) = \true$.
    By Lemma \ref{lem-prv}, $\nz(\data(C_k)) \subseteq [k-1]$. This shows that $\A(C_k) = \bigcup_{i \in \nz(\data(C_k))} \A(C_i)$ by Lemma \ref{lem-asm-for-lin-rnd} as $\formula_{ASM}(k) = \true$. It remains to show that the weighted linear combination $C$ is roundable, and that its rounding dominates $C_k$.
    Since $\formula_{RND}(A,B,eq) = \true$, we know $eq = \false$, so that exactly one of $geq$ and $leq$ is $\true$ by construction. This shows that $\data(C_k)$ generates a suitable linear combination of $C_1, \dots, C_d$ by definition of $geq$ and $leq$ (see Definition \ref{def-lin-constr-predicate}). Since $\formula_{RND}(A,B,eq) = \true$, this linear combination $C \triangleq \sum_{i \in \nz(\data(C_k))} \data(C_k)_i C_i$ is roundable. 
    
    It remains to show that $\rnd(C)$ dominates $C_k$. First note that $geq = (s(C) = 1)$ and $leq = (s(C) = -1)$ since $eq = \false$. We have two cases.
    \begin{itemize}
        \item $\bigAnd_{j \in [n]} (A_j = 0) \AND (\ttif geq \ttthen B > 0 \ttelse \ttif leq \ttthen
        B < 0 \ttelse \false) = \true$. This implies that $l(C) = l(\rnd(C)) = 0$. If $geq = \true$, then $s(C) = 1$ and $B > 0$, so $\ceil{B} > 0$; in this case, $\rnd(C)$ has the form $0 \ge \ceil{B}$, which is an absurdity, so that $\rnd(C)$ dominates $C_k$. Otherwise, $s(C) = -1$ and $B < 0$, so $\floor{B} < 0$; in this case, $\rnd(C)$ has the form $0 \le \floor{B}$, which is an absurdity, so that $\rnd(C)$ dominates $C_k$.
        
        \item $\bigAnd_{j \in [n]} (A_j = a_{k,j}) \AND (\ttif s(C_k) = 0 \ttthen \false \ttelse \ttif s(C_k) = 1 \ttthen (geq \AND (\ceil{B} \ge b_k)) \ttelse (leq \AND (\floor{B} \le b_k))) = \true$. Then $l(\rnd(C)) = l(C) = l(C_k)$. We have three subcases. If $s(C_k) = 0$, then we immediately get an evaluation of $\false$, a contradiction. If $s(C_k) = 1$, then $geq = \true$ and $\ceil{B} \ge b_k$, which immediately implies that $\rnd(C)$ dominates $C_k$ since $geq = \true$ is equivalent to $s(C) = 1 = s(C_k)$. Lastly, if $s(C_k) = -1$, then $leq = \true$ and $\floor{B} \le b_k$, which implies that $\rnd(C)$ dominates $C_k$ since $leq = \true$ is equivalent to $s(C) = -1 = s(C_k)$.
    \end{itemize}

\end{proof}

\begin{proof}[Proof of Lemma \ref{lem-disjunction}]
    The proof follows from Definitions \ref{def-disjunction} and \ref{def-disjunction-predicate}, after observing that $s(C_i) \ne 0 \AND s(C_i) + s(C_j) = 0$ holds if and only if $\{s(C_i), s(C_j)\} = \{1, -1\}$.
\end{proof}

\begin{proof}[Proof of Lemma \ref{lem-uns-valid}]
    Suppose $\reason(C_k) = \uns$ and all previous derived constraints are valid. Write $\data(C_k) = (i_1, l_1, i_2, l_2) \in [d]^4$.
    Suppose $C_k$ is valid. Also $\A(C_k) = (\A(C_{i_1}) \setminus \{l_1\}) \cup (\A(C_{i_2}) \setminus \{l_2\})$. It follows that $\A(C_k) \subseteq [k-1]$ since all constraints $C_{m+1}, \dots, C_{k-1}$ are valid. Hence $A_k^j = \false$ for all $j \in S_{> k}$. For all $j \in S_{< k}$, observe that $j \in \A(C_k)$ if and only if $j \in \A(C_{i_1})$ and $j \ne l_1$, or $j \in \A(C_{i_2})$ and $j \ne l_2$, so $A_k^j = (A_{i_1}^j \AND (j \ne l_1)) \OR (A_{i_2}^j \AND (j \ne l_2))$. Hence $\formula_{ASM}(k) = \true$. We also have $i_1 < k$ and $i_2 < k$ and $l_1 < k$ and $l_2 < k$, since $\data(C_k) \in [k-1]^4$. Since $C_{i_1}$ and $C_{i_2}$ both dominate $C_k$, Lemma \ref{lem-domination} shows that $\formula_{DOM}(C_{i_1}, C_k) = \true$ and $\formula_{DOM}(C_{i_2}, C_k) = \true$. Finally, we have $\formula_{DIS}(i_1, i_2) = \true$ by Definition \ref{def-valid-der-constraint} and Lemma \ref{lem-disjunction}. Hence $\formula_{DER}(k) = \true$.

    Conversely, suppose $\formula_{DER}(k) = \true$. Then $\data(C_k) \in [k-1]$ since $k > i_1$ and $k > i_2$ and $k > l_1$ and $k > l_2$. 
    Since $\formula_{DOM}(C_{i_1}, C_k) \AND \formula_{DOM}(C_{i_2}, C_k) = \true$, Lemma \ref{lem-domination} implies that $C_{i_1}$ dominates $C_k$ and $C_{i_2}$ dominates $C_k$. Since $\formula_{DIS}(l_1, l_2) = \true$, Lemma \ref{lem-disjunction} implies that $C_{l_1}$ and $C_{l_2}$ form a disjunction. Observe that $\A(C_k) \subseteq S_{< k}$ since $A_k^j = \false$ for $j \in S_{> k}$, from $\formula_{ASM}(k)$. If $j \in S_{< k}$, then the logical equality $A_k^j = (A_{i_1}^j \AND (j \ne l_1)) \OR (A_{i_2}^j \AND (j \ne l_2))$ is equivalent to $j \in \A(C_k) \Leftrightarrow j \in (\A(C_{i_1}) \setminus \{l_1\}) \cup (\A(C_{i_2}) \setminus \{l_2\})$, so that $\A(C_k) \cap S_{< k} = ((\A(C_{i_1}) \setminus \{l_1\}) \cup (\A(C_{i_2}) \setminus \{l_2\})) \cap S_{< k}$.
    Since $C_{m+1}, \dots, C_{k-1}$ are valid, a simple induction argument shows that $\A(C_i) \subseteq [i]$ for all $i \in [k-1]$. 
    Thus, since $i_1, i_2 \in [k-1]$, we have $(\A(C_{i_1}) \setminus \{l_1\}) \cup (\A(C_{i_2}) \setminus \{l_2\}) \subseteq [k-1]$. Then, the above assertion immediately implies that $\A(C_k) = (\A(C_{i_1}) \setminus \{l_1\}) \cup (\A(C_{i_2}) \setminus \{l_2\})$. Hence $C_k$ is valid.
\end{proof}

\begin{proof}[Proof of Lemma \ref{lem-sol-constr-valid}]
    The equivalence follows immediately from Definition \ref{def-valid-der-constraint} and Lemma \ref{lem-domination}.
\end{proof}

\begin{proof}[Proof of Lemma \ref{lem-der-valid}]
    Observe that every derived constraint $C_{m+1}, \dots, C_d$ is valid if and only if $\bigAnd_{m+1 \le k \le d} \formula_{DER}(k) = \true$, by Corollary \ref{cor-der}. Observe that the constraint $C_d$ is an absurdity if and only if $C_d$ dominates $0 \ge 1$. Furthermore, since $\A(C_d) \subseteq S$, we have $\A(C_d) = \emptyset$ if and only if $\bigAnd_{j \in S} (\neg A_d^j) = \true$. It is then easy to see from Definition \ref{def-der-valid} and Lemma \ref{lem-domination} that $\cert$ is DER-valid if and only if $\formula_{DER} = \true$.
\end{proof}

\begin{proof}[Proof of Theorem \ref{thm-equivalence}]
    This follows from Lemma \ref{lem-sol} and \ref{lem-der-valid}.
\end{proof}
\newpage
\section{Forged and Manipulated VIPR Files}
\label{App-ForgedManipuated}

In this section, we present the manipulated VIPR files mentioned in Sec.~\ref{sec:detectionForgedManipulated}. They are also available in \cite{GitHub-Mendes}.

\subsection{\texttt{forged1.vipr}}

\begin{Verbatim}[frame=single]
VER 1.0
VAR 2
x y
INT 2
0 1
OBJ max
2  0 1  1 1
CON 4 4
C1 L 1 1 0 1
C2 L 1 1 1 1
B1 G 0 1 0 1
B2 G 0 1 1 1
RTP range 1 1
SOL 1
opt 1  1 1
DER 1
C3 L 1 OBJ { sol } -1
\end{Verbatim}

\subsection{\texttt{forged2.vipr}}

\begin{Verbatim}[frame=single]
VER 1.0
VAR 2
x y
INT 2
0 1
OBJ max
2  0 1  1 1
CON 4 4
C1 L 1/2 1 0 1
C2 L 1/2 1 1 1
B1 G 1/3 1 0 1
B2 G 1/3 1 1 1
RTP range -inf 0
SOL 0
DER 0
\end{Verbatim}

\subsection{\texttt{manipulated1.vipr}}

\begin{Verbatim}[frame=single]
VER 1.0
VAR 2
x y
INT 2
0 1
OBJ min
0
CON 3 0
C1  G 1  2  0 2  1 3
C2  L 2  2  0 3  1 -4
C3  L 3  2  0 -1  1 6
RTP infeas
SOL 0
DER 11
A1  L 0   1  0 1  { asm } -1
A2  G 1   1  0 1  { asm } -1
A3  L 0   1  1 1  { asm } -1
C4  G 1   0       { lin 3  0 1  3 -2  5 -3 } 12
A4  G 1   1  1 1  { asm } -1
C5  G 1   0       { lin 3  2 -1/3  3 -1/3  7 2 } 12
C6  G 1/4 1  1 1  { lin 2  1 -1/4  4 3/4 } 10
C7  G 1   1  1 1  { rnd 1  9 1 } 11
C8  G 1   0       { lin 3 1 -1/3  2 -1  10 14/3 } 11
C9  G 1   0       { uns 6 5  8 7 } 13
C10 G 1   0       { uns 11 4  12 3 } -1
\end{Verbatim}

\end{document}